\documentclass[aoas,nameyear,MSNbibl,seceqn,noautosecdot,dvips]{arximspdf}
\usepackage{dcolumn}
\usepackage{graphicx}

\doi{10.1214/10-AOAS446}
\volume{5}
\issue{2B}
\pubyear{2011}
\firstpage{1456}
\lastpage{1487}

\makeatletter
\newcommand{\var}{\operatorname{var}}
\newcommand{\trace}{\operatorname{trace}}
\newcommand{\Normal}{\operatorname{Normal}}
\newcommand{\trans}{^{\mathrm{T}}}
\newcommand{\tY}{\widetilde{Y}}
\newcommand{\tQ}{\widetilde{Q}}
\newcommand{\tW}{\widetilde{W}}
\newcommand{\tU}{\widetilde{U}}
\newcommand{\tT}{\widetilde{T}}
\let\epsilon\varepsilon
\newcommand{\tepsilon}{\widetilde{\epsilon}}
\newcommand{\sumi}{\sum_{i=1}^n}
\newcommand{\sumk}{\sum_{k=1}^{m_i}}
\newcommand{\sumjJ}{\sum_{j=1}^{J}}
\newcommand{\sumell}{\sum_{\ell=1}^6}
\newcommand{\IW}{\mathrm{IW}}
\newcommand{\rest}{\mathrm{rest}}
\newcommand{\density}{\mathrm{density}}
\newcolumntype{d}[1]{D{.}{.}{#1}}

  \let\sv@tabnotetext\tabnotetext
  \let\sv@tabnotemark@fmt\tabnotemark@fmt
   \long\def\legend#1{{\let\tabnote@indent\leavevmode\sv@tabnotetext[]{}{#1}}}

\def\bptnote#1{}

\makeatother

\begin{document}
\begin{frontmatter}

\title{A new multivariate measurement error model with zero-inflated
dietary data, and its application to dietary assessment\thanksref{TITL1}}
\runtitle{A new multivariate model for dietary data}
\thankstext{TITL1}{Supported by National Science Foundation
Instrumentation Grant number 0922866.}

\begin{aug}
\author[A]{\fnms{Saijuan} \snm{Zhang}\thanksref{aut1,aut2}\ead[label=e1]{saijuan.zhang@merck.com}},
\author[B]{\fnms{Douglas} \snm{Midthune}\ead[label=e2]{midthund@mail.nih.gov}},
\author[C]{\fnms{Patricia M.}~\snm{Guenther}\ead[label=e3]{Patricia.Guenther@cnpp.usda.gov}},
\author[D]{\fnms{Susan M.} \snm{Krebs-Smith}\ead[label=e4]{krebssms@mail.nih.gov}},
\author[B]{\fnms{Victor} \snm{Kipnis}\ead[label=e5]{kipnisv@mail.nih.gov}},
\author[B]{\fnms{Kevin W.} \snm{Dodd}\ead[label=e6]{doddk@mail.nih.gov}},
\author[E]{\fnms{Dennis~W.}~\snm{Buckman}\ead[label=e7]{BuckmanD@imsweb.com}},
\author[F]{\fnms{Janet A.} \snm{Tooze}\ead[label=e8]{jtooze@wfubmc.edu}},
\author[G]{\fnms{Laurence} \snm{Freedman}\ead[label=e9]{lsf@actcom.co.il}}
\and
\author[H]{\fnms{Raymond J.} \snm{Carroll}\corref{}\thanksref{aut1}\ead[label=e10]{carroll@stat.tamu.edu}}
\thankstext{aut1}{Supported by a Grant from the National
Cancer Institute (CA57030).}
\thankstext{aut2}{This paper forms part of Zhang's Ph.D. dissertation at
Texas A\&M University.}
\runauthor{S. Zhang et al.}
\affiliation{}
\affiliation{Texas A\&M University, National Cancer Institute, U.S.
Department of Agriculture, National Cancer Institute, National Cancer
Institute, National~Cancer Institute,
Information Management Services, Inc., Wake~Forest~University, Sheba
Medical Center and Texas A\&M University}
\address[A]{S. Zhang\\Merck \& Co., Inc.\\
126 E. Lincoln Ave.\\ PO Box 2000 (RY34-A316)\\
Rahway, New Jersey 07065\\USA\\
\printead{e1}}
\address[B]{D. Midthune\\
V. Kipsnis\\
K. Dodd\\
Biometry Research Group\\
Division of Cancer Prevention\\
National Cancer Institute\\
6130 Executive Boulevard EPN-3131\\
Bethesda, Maryland 20892-7354\\
USA\\
\printead{e2}\\
\hphantom{E-mail: }\printead*{e5}\\
\hphantom{E-mail: }\printead*{e6}}
\address[C]{
P. M. Guenther\\
Center for Nutrition Policy and Promotion\\
U.S. Department of Agriculture\\
3101 Park Center Drive, Ste. 1034\\
Alexandria, Virginia 22302\\
USA\\
\printead{e3}}
\address[D]{S. Krebs-Smith\\
Applied Research Program\\
Division of Cancer Control\\
\quad and Population Sciences\\
National Cancer Institute\\
6130 Executive Boulevard, EPN-4005\\
Bethesda, Maryland 20892\\
USA\\
\printead{e4}}
\address[E]{D. Buckman\\
Information Management Services, Inc.\\
12501 Prosperity Drive\\
Silver Spring, Maryland 20904\\
USA\\
\printead{e7}}
\address[F]{J. Tooze\\
Department of Biostatistical Sciences\\
Wake Forest University, School \\
\quad of Medicine\\
Medical Center Boulevard\\
Winston-Salem, North Carolina 27157\\
USA\\
\printead{e8}}
\address[G]{L. Freedman\\
Gertner Institute for Epidemiology\\
\quad and Health Policy Research\\
Sheba Medical Center\\
Tel Hashomer 52161\\
Israel\\
\printead{e9}}
\address[H]{R. J. Carroll\\
Department of Statistics\\
Texas A\&M University\\
3143 TAMU\\
College Station, Texas 77843-3143\\
USA\\
\printead{e10}}
\end{aug}

% HISTORY:
\received{\smonth{4} \syear{2010}}
\revised{\smonth{11} \syear{2010}}

% ABSTRACT
%
\begin{abstract}
In the United States the preferred method of obtaining dietary intake
data is the 24-hour dietary recall, yet the measure of most interest is
usual or long-term average daily intake, which is impossible to
measure. Thus, usual dietary intake is assessed with considerable
measurement error. Also, diet represents numerous foods, nutrients and
other components, each of which have distinctive attributes. Sometimes,
it is useful to examine intake of these components separately, but
increasingly nutritionists are interested in exploring them
collectively to capture overall dietary patterns. Consumption of these
components varies widely: some are consumed daily by almost everyone on
every day, while others are episodically consumed so that 24-hour
recall data are zero-inflated. In addition, they are often correlated
with each other. Finally, it is often preferable to analyze the amount
of a dietary component relative to the amount of energy (calories) in a
diet because dietary recommendations often vary with energy level. The
quest to understand overall dietary patterns of usual intake has to
this point reached a standstill. There are no statistical methods or
models available to model such complex multivariate data with its
measurement error and zero inflation. This paper proposes the first
such model, and it proposes the first workable solution to fit such a
model. After describing the model, we use survey-weighted MCMC
computations to fit the model, with uncertainty estimation coming from
balanced repeated replication. The methodology is illustrated through
an application to estimating the population distribution of the Healthy
Eating Index-2005 (HEI-2005), a multi-component dietary quality index
involving ratios of interrelated dietary components to energy, among
children aged 2--8 in the United States. We pose a number of interesting
questions about the HEI-2005 and provide answers that were not
previously within the realm of possibility, and we indicate ways that
our approach can be used to answer other questions of importance to
nutritional science and public health.
\end{abstract}

% KEYWORDS
%
\begin{keyword}
\kwd{Bayesian methods}
\kwd{dietary assessment}
\kwd{Latent variables}
\kwd{measurement error}
\kwd{mixed models}
\kwd{nutritional epidemiology}
\kwd{nutritional surveillance}
\kwd{zero-inflated data}.
\end{keyword}

\end{frontmatter}

%s1 ###
\section{\texorpdfstring{Introduction.}{Introduction}}\label{intro}

This paper presents statistical models and methodology to overcome a
major stumbling block in the field of dietary assessment.
More nutritional background is provided in Section \ref{sec:sec2}: a
summary of the key conceptual issues follows:
\begin{itemize}
\item Nutritional surveys conducted in the United States typically use
24-hour (24 h) dietary recalls to obtain intake data, that is, an
assessment of what was consumed in the past 24 hours.
\item Because dietary recommendations are intended to be met over time,
nutritionists are interested in ``usual'' or long-term average daily intake.
\item Dietary intake is thus assessed with considerable measurement error.
\item Consumption patterns of dietary components vary widely; some are
consumed daily by almost everyone, while others are episodically
consumed so that 24-hour recall data are zero-inflated. Further, these
components are correlated with one another.
\item Nutritionists are interested in dietary components collectively
to capture patterns of \textit{usual} dietary intake, and thus need
multivariate models for usual intake.
\item These multivariate models for usual intakes, taking into account
episodically consumed foods, do not exist, nor do methods exist for
fitting them.
\end{itemize}

One way to capture dietary patterns is by scores, although our work is
not limited to scores. The Healthy Eating Index-2005 (HEI-2005),
described in detail in Section \ref{sec:sec2}, is a scoring system
based on a priori knowledge of dietary recommendations, and is on a
scale of 0--100. Ideally, it consists of the \textit{usual}
intake of 6 episodically consumed and thus 24 h-zero inflated foods, 6
daily-consumed dietary components, adjusts these for energy (caloric)
intake, and gives a score to each component. The total score is the sum
of the individual component scores. Higher scores indicate greater
compliance with dietary guidelines and, therefore, a healthier diet.
Here are a few questions that nutritionists have not been able to
answer, and that our approach can address:
\begin{itemize}
\item What is the distribution of the HEI-2005 total score, and what \%
of Americans are eating a healthier diet defined, for example, by a
total score exceeding 80?
\item What is the correlation between the individual score on each
dietary component and the scores of all other dietary components?
\item Among those whose total HEI-2005 score is $> $50 or $\leq$50,
what is the distribution of usual intake of whole grains, whole fruits,
dark green and orange vegetables and legumes (DOL) and calories from
solid fats, alcoholic beverages and added sugars (SoFAAS)?
\item What \% of Americans exceed the median score on all 12 HEI-2005
components?
\end{itemize}

In this paper, to answer public health questions such as these that can
have policy implications, we build a novel multivariate measurement
error model for estimating the distributions of usual intakes, one that
accounts for measurement error and zero-inflation, and has a special
structure associated with the zero-inflation. Previous attempts to fit
even simple versions of this model, using nonlinear mixed effects
software, failed because of the complexity and dimensionality of the
model. We use survey-weighted Monte Carlo computations to fit the model
with uncertainty estimation coming from balanced repeated replication.
The methodology is illustrated using the HEI-2005 to assess the diets
of children aged 2--8 in the United States. This work represents the
first analysis of joint distributions of usual intakes for multiple
food groups and nutrients.

The paper is outlined as follows. In Section \ref{sec:sec2} we give
the background for the data we observe. In particular, we provide more
information about the HEI-2005. Section \ref{sec:sec3} describes our
model which is a highly nonlinear, zero-inflated, repeated measures
model with multiple latent variables. The model also has a patterned
covariance matrix with structural zeros and ones. We derive a
parameterization that allows estimated covariance matrices to be actual
covariance matrices. We also define technically what we mean by usual
intake, and illustrate the use of simulation methods used to answer the
questions posed above, as well as many others.

Section \ref{sec:howtoestimate} describes our estimation procedure.
Previous attempts using nonlinear mixed effects models to estimate the
distribution of episodically consumed food groups [Tooze et~al.
(\citeyear{Tooetal06}); \citet{Kipetal09}] do not work here because of the high
dimensionality of the problem. We instead develop a~Monte Carlo
strategy based on the idea of Gibbs sampling; although because of
sampling weights, we treat the method as a frequentist (non-Bayesian)
one. This section describes some of the basics of the methodology; the
full technical details of implementation are given in the \hyperref[app]{Appendix}.

Section \ref{sec:sec5} describes the analysis of the HEI-2005
components using the 2001--2004 National Health and Nutrition
Examination Survey (NHANES) for children ages 2--8. Important contextual
points arise because of the nature of the data. For example, if whole
grains are consumed, then necessarily total grains are consumed with
probability one, a restriction that a naive use of our model cannot
handle. We develop a simple novel device to uncouple consumption
variables that are tightly linked in this way. Finally, in this section
we provide the first answers to the four questions we have posed. In
Section \ref{sec:sec6} we discuss various additional aspects of the
problem and the data analysis. Concluding remarks and a policy
application are given in Section~\ref{sec:sec7}.

There are a number of general reviews of the measurement error field
[\citet{Ful87}; Gustafson (\citeyear{Gus04});
\citet{Caretal06}; \citet{Buo10}]. Recent papers that focus on estimating the density function of
a univariate continuous random variable subject to measurement error
include \citet{Del08}, Delaigle and Hall (\citeyear{DelHal08}, \citeyear{DelHal10}), \citet{DelMei08},
\citet{DelHalMei08}, \citet{StaRupBuo08}
and \citet{Wan98}. The field of measurement error in regression continues
to expand rapidly, with some recent contributions including \citet{KucMwaLes06}, \citet{Guo08}, \citet{Liaetal08}, \citet{MesNat08} and \citet{Nat09}. There is also a large
statistical literature on measurement error as it relates to public
health nutrition: some recent papers relevant to our work include
Carriquiry (\citeyear{Car99}, \citeyear{Car03}), \citet{Feretal09}, \citet{FraSha04}, \citet{Kotetal09},
Nusser et~al. (\citeyear{Nusetal96}), \citet{NusFulGue97}, Prentice (\citeyear{Pre96}, \citeyear{Pre03}), \citet{TooGruJon02} and
Tooze et~al. (\citeyear{Tooetal06}).

%s2 ###
\section{\texorpdfstring{Data and the HEI-2005 scores.}{Data and the HEI-2005 scores}}\label{sec:sec2}

Here we give more detail about the nutrition context that motivates
this work.

%t1 ###
\begin{table}[b]
\caption{Description of the HEI-2005 scoring system. Except for saturated fat
and SoFAAS, density is obtained by multiplying usual intake by 1000 and
dividing by usual intake of kilocalories}\label{tab:heidescription}
\begin{tabular*}{\tablewidth}{@{\extracolsep{\fill}}lll@{}}
\hline
\textbf{Component} & \textbf{Units} & \textbf{HEI-2005 score calculation} \\
\hline
Total fruit & cups & $\min{(5, 5\times(\density/0.8))}$ \\
Whole fruit & cups & $\min{(5, 5\times(\density/0.4))}$\\
Total vegetables & cups & $\min{(5, 5\times(\density/1.1))}$\\
DOL & cups & $\min{(5, 5\times(\density/0.4))}$\\
Total grains & ounces & $\min{(5, 5\times(\density/3))}$\\
Whole grains & ounces & $\min{(5, 5\times(\density/1.5))}$\\
Milk & cups & $\min{(10, 10\times(\density/1.3))}$\\
Meat and beans & ounces & $\min{(10, 10\times(\density/2.5))}$\\
Oil & grams & $\min{(10, 10\times(\density/12))}$\\
Saturated fat & \% of & if $\density\ge15$ score${}={}$0\\
& energy & else if $\density\le7$ score${}={}$10\\
& & else if $\density>10$ $\mbox{score} = 8- (8\times(\density- 10)/5)$\\
& & else, $\mbox{score} = 10 - (2\times(\density- 7)/3)$\\
Sodium & milligrams & if $\density\ge2000$ score${}={}$0\\
& & else if $\density\le700$ score${}={}$10\\
& & else if $\density\ge1100$ \\
& & $ \mbox{score} = 8 - \{8\times( \density- 1100)/(2000 -
1100)\}$\\
& & else $ \mbox{score} = 10 - \{2\times(\density- 700)/(1100 - 700)\}$\\
SoFAAS & \% of & if $\density\ge50$ score${}={}$0\\
& energy & else if $\density\le20$ score${}={}$20\\
& & else $\mbox{score} = 20 - \{20\times(\density- 20)/(50 - 20)\}$\\
\hline
\end{tabular*}
\legend{For saturated fat, density
is $9 \times100 $ usual saturated fat (grams) divided by usual
calories, that is, the percentage of usual calories coming from usual
saturated fat intake. For SoFAAS, the density is the percentage of
usual intake that comes from usual intake of calories, that is, the
division of usual intake of SoFAAS by usual intake of calories. Here,
``DOL'' is dark green and orange vegetables and legumes. Also,
``SoFAAS'' is calories from solid fats, alcoholic beverages and added
sugars. The total HEI-2005 score is the sum of the individual component scores.}\vspace*{-1pt}
\end{table}

In surveys conducted in the United States, the preferred method of
obtaining intake data is the 24-hour dietary recall because it limits
respondent burden and facilitates accurate reporting; yet the measure
of greatest interest is ``usual'' or long-term average daily intake.
Thus, dietary intake is assessed with considerable measurement error.
Also, diets are comprised of numerous foods, nutrients and other
components, each of which may have distinctive attributes and effects
on nutritional health. Sometimes, it is useful to examine intake of
these components separately, but increasingly nutritionists are
interested in exploring them collectively to capture patterns of
dietary intake. Consumption patterns of these components vary widely;
some are consumed daily by almost everyone, while others are
episodically consumed so that 24-hour recall data are zero-inflated. In
addition, these various components are often correlated with one
another. Finally, it is often preferable to analyze the amount of a
dietary component relative to the amount of energy (calories) in a diet
because dietary recommendations often vary with energy level, and this
approach provides a way of standardizing dietary assessments.

One of the US Department of Agriculture's (USDA's) strategic objectives
is ``to promote healthy diets'' and it has developed an associated
performance measure, the Healthy Eating Index-2005 (HEI-2005,
\href{http://www.cnpp.usda.gov/HealthyEatingIndex.htm}{http://}
\href{http://www.cnpp.usda.gov/HealthyEatingIndex.htm}{www.cnpp.usda.gov/HealthyEatingIndex.htm}).
The HEI-2005 is based
on the key recommendations of the 2005 Dietary Guidelines for
Americans
(\href{http://www.health.gov/dietaryguidelines/dga2005/document/default.htm}%
{http://www.health.gov/dietaryguidelines/dga2005/document/default.htm}).
The index includes ratios of interrelated dietary components to energy.
The HEI-2005 comprises 12 distinct component scores and a total summary
score. See Table \ref{tab:heidescription} for a list of these components and the standards for
scoring, and see \citet{GueReeKre08} for details. Intakes of each
food or nutrient, represented by one of the 12 components, are
expressed as a ratio to energy intake, assessed, and ascribed a score.

The HEI-2005 is used to evaluate the diets of Americans to assess
compliance with the 2005 Dietary Guidelines, yet use of the HEI-2005 is
limited by the challenges described above. Until recently, there have
been no solutions to these challenges, so published evaluations have
been limited to analyses of mean scores for the population and various
subgroups. \citet{Freetal10} have described a method of
estimating the population distribution of a single component of
HEI-2005, and the prevalence of high or low scores on that component;
but there has been to date no satisfactory way to determine the
prevalence of high or low total HEI-2005 scores, considering all of its
interrelated components simultaneously. In addition, answers to the
complex questions posed in the \hyperref[intro]{Introduction} remain unavailable. This
paper aims to provide a means to do these crucial evaluations.

The 12 HEI-2005 components represent 6 episodically consumed food
groups (total fruit, whole fruit, total vegetables,\vadjust{\goodbreak} dark green and
orange vegetables and legumes or DOL, whole grains and milk), 3
daily-consumed food groups (total grains, meat and beans and oils) and
3 other daily-consumed dietary components (saturated fat; sodium; and
calories from solid fats, alcoholic beverages and added sugars, or
SoFAAS). The classification of food groups as ``episodically'' and
``daily'' consumed is based on the number of individuals who report
them on  24~h recalls. If there are only a few zeros for a component, we
treat that as a daily-consumed food, and replace all zeros with $1/2$ the
minimum value of the nonzeros for that food. However, the crucial
statistical aspect of the data is that six of the food groups are
zero-inflated. The percentages of reported nonconsumption of total
fruit, whole fruit, whole grains, total vegetables, DOL and milk on any
single day are 17\%, 40\%, 42\%, 3\%, 50\% and 12\%, respectively.

We are interested in the usual intake of foods for children aged 2--8.
The data available to us, described in more detail in Section \ref
{sec:sec5}, came from the National Health and Nutrition Examination
Survey, 2001--2004 (NHANES). The data used here consisted of $n =2\mbox{,}638$
children, each of whom had a~survey weight $w_i$ for $i=1,\ldots,n$. In
addition, one or two  24 h dietary recalls were available for each
individual. Along with the dietary variables, there are covariates such
as age, gender, ethnicity, family income and dummy variables that
indicate a weekday or a weekend day, and whether the recall was the
first or second reported for that individual.

Using the 24 h recall data reported, for each of the episodically
consumed food groups, two variables are defined: (a) whether a food
from that group was consumed; and (b) the amount of the food that was
reported on the  24 h recall. For the 6 daily-consumed food groups and
nutrients, only one variable indicating the consumption amount is
defined. In addition, the amount of energy that is calculated from the
 24 h recall is of interest. The number of dietary variables for each
 24 h recall is thus $12+6+1 = 19$. The observed data are $Y_{ijk}$ for
the $i$th person, the $j$th variable and the $k$th replicate,
$j=1,\ldots,19$ and $k=1,\ldots,m_i$. In the data set, at most two
 24 h recalls were observed, so that $m_i \leq2$. Set $\tY_{ik} =
(Y_{i1k},\ldots,Y_{i,19,k})\trans$, where
\begin{itemize}
\item$Y_{i, 2\ell-1, k} = \mbox{Indicator}$ of whether dietary component \#$\ell$ is consumed, with $\ell= 1, 2, 3, 4, 5, 6$.
\item$Y_{i, 2\ell,k} = \mbox{Amount}$ of food \#$\ell$ consumed. This
equals zero, of course, if none of food \#$\ell$ is consumed,
with $\ell= 1, 2, 3, 4, 5, 6$.
\item$Y_{i,\ell+6,k} = \mbox{Amount}$ of nonepisodically consumed food or
nutrient \#$\ell$,
with $\ell= 7, 8, 9, 10, 11, 12$.
\item$Y_{i,19,k} = \mbox{Amount}$ of energy consumed as reported by the  24 h recall.
\end{itemize}

%s3 ###
\section{\texorpdfstring{Model and methods.}{Model and methods}}\label{sec:sec3}
%s3.1 ###
\subsection{\texorpdfstring{Basic model description.}{Basic model description}} \label{sec:sec2_1}

Our model is a generalization of work by \citet{Tooetal06} and
\citet{Kipetal09} for a single food and Kipnis et~al. (\citeyear{Kipetal}) and
\citet{SMi} for a single food and nutrient. Observed data will
be denoted as $Y$, and covariates in the model will be denoted as $X$.
As is usual in measurement error problems, there will also be latent
variables, which will be denoted by $W$.

We use a probit threshold model. Each of the 6 episodically consumed
foods will have 2 sets of latent variables, one for consumption and one
for amount, while the 6 daily-consumed foods and nutrients as well as
energy will have 1 set of latent variables, for a total of 19. The
latent random variables are $\epsilon_{ijk}$ and $U_{ij}$, where
$(U_{i1}, \ldots, U_{i,19}) = \Normal(0,\Sigma_u)$ and $(\epsilon
_{i1k},\ldots,\epsilon_{i,19,k}) = \Normal(0,\Sigma_{\epsilon})$
are mutually independent. In this model, food $\ell= 1,\ldots,6$ being
consumed on day $k$ is equivalent to observing the binary $Y_{i,2\ell
-1, k}$, where
\begin{eqnarray} \label{eq:qkipnis03}
&&Y_{i,2\ell-1, k} = 1 \nonumber\\[-8pt]\\[-8pt]
&&\qquad  \iff\quad  W_{i,2\ell-1, k} = X_{i,2\ell-1,k}\trans\beta_{2\ell-1} + U_{i,2\ell-1} +
\epsilon_{i,2\ell-1, k} > 0.\nonumber
\end{eqnarray}
If the food is consumed, we model the amount reported $Y_{i, 2\ell,
k}$ as
%
%e3.1 ###
\begin{eqnarray} \label{eq:qkipnis04}
[g_{\mathrm{tr}}(Y_{i,2\ell, k},\lambda_\ell) \vert Y_{i,2\ell-1, k}=1 ]
&=& W_{i,2\ell, k} \nonumber\\[-8pt]\\[-8pt] \nonumber
&=& X_{i,2\ell,k}\trans\beta_{2\ell} + U_{i,2\ell} + \epsilon
_{i,2\ell, k},
\end{eqnarray}
where $g_{\mathrm{tr}}(y,\lambda) = \sqrt{2} \{g(y,\lambda) - \mu
(\lambda)\}/\sigma(\lambda)$, $g(y,\lambda)$ is the usual Box-Cox
transformation with transformation parameter $\lambda$, and $\{\mu
(\lambda),\sigma(\lambda)\}$ are the sample mean and standard
deviation of $g(y,\lambda)$, computed from the nonzero food data. This
standardization is simply a convenient device to improve the numerical
performance of our algorithm without affecting the conclusions of our analysis.

The reported consumption of daily consumed foods or nutrients $\ell=\break
7, \ldots, 12$ is modeled as
%
%e3.2 ###
\begin{eqnarray}\label{eq:qkipnis05a}
g_{\mathrm{tr}}(Y_{i, \ell+6, k},\lambda_\ell)&=& W_{i,\ell+6, k} =
X_{i,\ell+6,k}\trans\beta_{\ell+6} + U_{i, \ell+6} + \epsilon
_{i,\ell+6, k}.
\end{eqnarray}
Finally, energy is modeled as
%
%e3.3 ###
\begin{eqnarray}\label{eq:qkipnis05b}
g_{\mathrm{tr}}(Y_{i, 19, k},\lambda_{13})&=& W_{i,19, k} =
X_{i,19,k}\trans\beta_{19} + U_{i, 19} + \epsilon_{i,19, k}.
\end{eqnarray}
As seen in (\ref{eq:qkipnis04})--(\ref{eq:qkipnis05b}), different
transformations $(\lambda_1,\ldots,\lambda_{13})$ are allowed to be used
for the different types of dietary components; see Section \ref
{esttransformation}.

In summary, there are latent variables $\tW_{ik} = (W_{i1k},\ldots,W_{i,
19, k})\trans$, latent random effects $\tU_i =
(U_{i1},\ldots,U_{i,19})\trans$, fixed effects $(\beta_1,\ldots,\beta
_{19})$, and design matrices $(X_{i1k},\ldots,X_{i,19,k})$. Define
$\tepsilon_{ik} = (\epsilon_{i1k},\ldots,\epsilon_{i, 19, k})\trans$.
The latent variable model is
%
%e3.4 ###
\begin{eqnarray}\label{eq:qe01}
W_{ijk} &=& X_{ijk}\trans\beta_j + U_{ij} + \epsilon_{ijk},
\end{eqnarray}
where $\tU_i = \Normal(0,\Sigma_u)$ and $\tepsilon_{ik} = \Normal
(0,\Sigma_{\epsilon})$ are mutually independent.

%s3.2 ###
\subsection{\texorpdfstring{Restriction on the covariance matrix.}{Restriction on the covariance matrix}}\label{secrest}
Two necessary restrictions are set on $\Sigma_{\epsilon}$. First,
following Kipnis et~al. (\citeyear{Kipetal09}, \citeyear{Kipetal}), $\epsilon_{i,2\ell-1, k}$ and
$\epsilon_{i, 2\ell, k}$, ($\ell=1,\ldots,6$) are set to be
independent. Second, in order to technically identify $\beta_{2\ell
-1}$ and the distribution of $U_{i, 2\ell-1}$ ($\ell=1,\ldots,6$),
we require that $\var(\epsilon_{i, 2\ell-1, k}) = 1$, because
otherwise the marginal probability of consump-\break tion of dietary component
$\#\ell$ would be $\Phi\{(X_{i,2\ell-1,k}\trans\beta_{2\ell-1} +
U_{i, 2\ell-1})/\break\var^{1/2}(\epsilon_{i, 2\ell-1, k})\}$, and thus
components of $\beta$ and $\Sigma_u$ would be identified only up to
the scale $\var^{1/2}(\epsilon_{i, 2\ell-1, k})$.

So that we can handle any number of episodically consumed dietary
components and any number of daily consumed components, suppose that
there are $J$ episodically consumed dietary components, and $K$ daily
consumed dietary components, and in addition there is energy. Then the
restrictions defined above lead to the covariance matrix
\begin{eqnarray}\label{eq:qszmc01}
\Sigma_{\epsilon}& =&
\pmatrix{
% 1 & 0 & s_{13} & s_{14} & \dots& s_{1,2J+1} & \dots& s_{1,2J+K+1}
% 0 & s_{22} & s_{23} & s_{24} & \dots& s_{2,2J+1} & \dots&
%s_{2,2J+K+1}\cr
% s_{13} & s_{23} & 1 & 0 & \dots& s_{3,2J+1} & \dots& s_{3,2J+K+1}\cr
% s_{14} & s_{24} & 0 & s_{44} & \dots& s_{4,2J+1} & \dots&
%s_{4,2J+K+1}\cr
% \vdots& \vdots& \vdots&\vdots& \ddots& \vdots& \dots& \vdots
% s_{1,2J+1} & s_{2,2J+1} & s_{3,2J+1}&s_{4,2J+1}& \dots&
%s_{2J+1,2J+1}& \dots& s_{2J+1,2J+K+1}\cr
% \vdots& \vdots& \vdots& \vdots& \vdots& \vdots&\ddots& \vdots
% s_{1,2J+K+1} &s_{2,2J+K+1} & s_{3,2J+K+1}&s_{4,2J+K+1}&\dots&
%s_{2J+1,2J+K+1} & \dots& s_{2J+K+1,2J+K+1}\cr
%%%%%%%%%%%%%%%%%%%%%%%%%%%%%%%%%%%%%%%%%%%%%%%%%%%%%%%%%%%%%%%%%%%%%%%%%%%%%%%%%%%%%%%%%%%%%%%%%%%%%%%%%%%%%%%%%%%%%%%%%%%%%%%%%%%%%%%%%%%%%%%%%%%%%%%%%
1 & 0 & s_{13} & s_{14} & \dots\cr
0 & s_{22} & s_{23} & s_{24} & \dots\cr
s_{13} & s_{23} & 1 & 0 & \dots\cr
s_{14} & s_{24} & 0 & s_{44} & \dots\cr
\vdots& \vdots& \vdots&\vdots& \ddots\cr
s_{1,2J+1} & s_{2,2J+1} & s_{3,2J+1}&s_{4,2J+1}& \dots\cr
\vdots& \vdots& \vdots& \vdots& \vdots\cr
s_{1,2J+K+1} &s_{2,2J+K+1} & s_{3,2J+K+1}&s_{4,2J+K+1}&\dots\cr
}
\nonumber\\[-8pt]\\[-8pt]
&&{}\times
\pmatrix{
\dots& s_{1,2J+1} & \dots& s_{1,2J+K+1} \cr
\dots& s_{2,2J+1} & \dots& s_{2,2J+K+1}\cr
\dots& s_{3,2J+1} & \dots& s_{3,2J+K+1}\cr
\dots& s_{4,2J+1} & \dots& s_{4,2J+K+1}\cr
\ddots& \vdots& \dots& \vdots\cr
\dots& s_{2J+1,2J+1}& \dots& s_{2J+1,2J+K+1}\cr
\vdots& \vdots&\ddots& \vdots\cr
\dots& s_{2J+1,2J+K+1} & \dots& s_{2J+K+1,2J+K+1}\cr
}.\nonumber
\end{eqnarray}

The difficulty with parameterizations of (\ref{eq:qszmc01}) is that
the cells that are not constrained to be $0$ or $1$ cannot be left
unconstrained, otherwise (\ref{eq:qszmc01}) need not be a covariance
matrix, that is, positive semidefinite.

We have developed an unconstrained parameterization that results in the
structure (\ref{eq:qszmc01}). Consider an unconstrained lower
triangular matrix $V$ and define $\Sigma_{\epsilon} = V V\trans$.
This is positive semidefinite and therefore qualifies $\Sigma
_{\epsilon}$ as a proper covariance matrix. The form of $V$ is
\begin{eqnarray*}
V =
\pmatrix{
v_{11} & 0 & \dots&0 \cr
v_{21} & v_{22} & \dots&0 \cr
\vdots& \vdots&\ddots& \vdots\cr
v_{2J+K+1,1} & v_{2J+K+1,2} & \dots& v_{2J+K+1,2J+K+1}\cr
}
.
\end{eqnarray*}
To achieve the desired pattern (\ref{eq:qszmc01}), we derive the
following four restrictions:
\begin{eqnarray*}
v_{11} &=& 1; \\
v_{21} &=& 0; \\
\sum_{p=1}^q v_{qp}^2 &=& 1;\qquad  q=3,5,\ldots,2J-1;\\
\sum_{p=1}^q v_{qp}v_{q+1,p} &=& 0;\qquad  q=3,5,\ldots,2J-1.
\end{eqnarray*}
The third restriction can be ensured by the further parameterization
\begin{eqnarray*}
v_{31} &=& r_1 \sin(\theta_{1});\\
v_{32} &=& r_1 \cos(\theta_{1}); \\
v_{33} &=& \sqrt{1 - r_1^2};\\
v_{2q+1,1} &=&r_q \sin\bigl(\theta_{1+(q-1)^2}\bigr);\\
v_{2q+1,p} &=& r_q \cos\bigl(\theta_{1+(q-1)^2}\bigr)\times\cdots\times\cos
\bigl(\theta_{p-1+(q-1)^2}\bigr)\sin\bigl(\theta_{p+(q-1)^2}\bigr), \\
&& p = 2, \ldots, 2q-1; \\
v_{2q+1,2q} &=& r_q \cos\bigl(\theta_{1+(q-1)^2}\bigr)\times\cdots\times\cos
(\theta_{q^2}) ; \\
v_{2q+1,2q+1} &=&\sqrt{1 - r_q^2},
\end{eqnarray*}
where $q = 2, 3, \ldots,J-1$, $\vert r_t \vert\leq1$, $t = 1, \ldots
, J-1$, and $\vert\theta_{s}\vert\leq\pi$, $s = 1, \ldots,\break (J-1)^2$.

Similarly, the fourth restriction can be further expressed by setting
\begin{eqnarray*}
v_{q+1,q} = -\sum_{p=1}^{q-1} v_{qp}v_{q+1,p} / v_{qq} = -\sum
_{p=1}^{q-1} v_{qp}v_{q+1,p} / \sqrt{1 - r_{(q-1)/2}^2},
\end{eqnarray*}
where $q = 3, 5, \ldots,2J-1$.

Note that $\vert\Sigma_{\epsilon} \vert= \vert V \vert^ 2 = \prod
_{q=1}^{2J+K+1} v_{qq}^2 = \prod_{q=1}^{J} v_{2q,2q}^2 \prod
_{q=2J+1}^{2J+K+1} v_{q,q}^2 \prod_{q=1}^{J-1}(1-r_q^2)$.

%s3.3 ###
\subsection{\texorpdfstring{The use of sampling weights.}{The use of sampling weights}}\label{weights}

As described in the \hyperref[app]{Appendix}, we used the survey sample weights from
NHANES both in the model fitting procedure and, after having fit the
model, in estimating the distributions of usual intake.

While not displayed here, we redid the model fitting calculations
without weighting, because the covariates we use are major players in
determining the sampling weights, hence, it is reasonable to believe
that the model in Section \ref{sec:sec3} holds both in the sample and
in the population. When we did this, the parameter estimates were
essentially unchanged.

Thus, we use the sampling weights only for estimation of the population
distributions. We actually did this for the purpose of handling the
clustering in the sample design. For such a complex statistical
procedure as ours, we knew we could not do theoretical standard errors,
so we thought about the bootstrap, and realized that putting together a
bootstrap for the complex survey would be nearly impossible. However,
we already had developed a~set of Balanced Repeated Replication (BRR)
weights [Wolter (\citeyear{Wol07})]; see Section \ref{sec5.5} for details. These
BRR weights have the property that, in the frequentist survey sampling
sense, they appropriately reflect the clustering in the standard error
calculations.

Of course, the use of sampling weights in the modeling provide unbiased
estimates of the (super) population parameters of interest. In
addition, the use of sampling weights in the distribution estimation
provides an estimated distribution that is representative of the US
population, not just the sample.

%s3.4 ###
\subsection{\texorpdfstring{Distribution of usual intake and the HEI-2005
scores.}{Distribution of usual intake and the HEI-2005
scores}}\label{subsec:dhei}

We assume here that estimates of $\Sigma_u$, $\Sigma_{\epsilon}$ and
$\beta_j$ for $j=1,\ldots,19$ have been constructed; see Section \ref
{sec:howtoestimate}. Here we discuss what we mean by usual intake for
an individual, how to estimate the distribution of usual intakes, how
to convert usual intakes into HEI-2005 scores, and how to assess uncertainty.

Consider the first episodically consumed dietary component, a food
group, with reporting being done on a weekend. Set $X_{i1,\mathrm{wkend}}$
and $X_{i2,\mathrm{wkend}}$ to be the versions of $X_{i1k}$ and $X_{i2k}$
where the dummy variable has the indicator of the weekend and that the
recall is the first one. Following \citet{Kipetal09}, we define the
usual intake for an individual on the weekend to be the expectation of
the reported intake conditional on the person's random effects $\tU
_i$. Let the $(q,p)$ element of $\Sigma_{\epsilon}$ be denoted as
$\Sigma_{\epsilon,q,p}$. As in Kipnis et~al. (\citeyear{Kipetal09}) define
%
%e3.5 ###
\begin{equation}
g_{\mathrm{tr}}^*\{v,\lambda,\Sigma_{\epsilon,q,p}\} = g_{\mathrm{tr}}^{-1}(v,\lambda) + \frac{1}{2}\Sigma_{\epsilon,q,p} \frac
{\partial^2 g_{\mathrm{tr}}^{-1}(v,\lambda)}{\partial v^2}.
\end{equation}
Detailed formulas for this are given in Appendix \ref
{backtransformation}. Then, following the convention of \citet{Kipetal09}, the person's usual intake of the first episodically consumed
dietary component on the weekend is defined as
\begin{eqnarray*}
T_{i1,\mathrm{wkend}} = \Phi(X_{i1,\mathrm{wkend}}\trans\beta_1 +
U_{i1})g_{\mathrm{tr}}^* (X_{i2,\mathrm{wkend}}\trans\beta_2 + U_{i2},\lambda
_1,\Sigma_{\epsilon,2,2} ).
\end{eqnarray*}
Similarly, let $X_{i1,\mathrm{wkday}}$ and $X_{i2,\mathrm{wkday}}$ be as above
but the dummy variable is appropriate for a weekday. Then the person's
usual intake of the first episodically consumed food group on weekdays
is defined as
\begin{eqnarray*}
T_{i1,\mathrm{wkday}} = \Phi(X_{i1,\mathrm{wkday}}\trans\beta_1 +
U_{i1})g_{\mathrm{tr}}^* (X_{i2,\mathrm{wkday}}\trans\beta_2 + U_{i2},\lambda
_1,\Sigma_{\epsilon,2,2} ).
\end{eqnarray*}
Finally, the usual intake of the first episodically consumed food for
the individual is
\begin{eqnarray*}
T_{i1} = (4 T_{i1,\mathrm{wkday}} + 3 T_{i1,\mathrm{wkend}} )/7,
\end{eqnarray*}
since Fridays, Saturdays and Sundays are considered to be weekend days.
Usual intake for the other episodically consumed food groups is defined
similarly.

A person's usual intake of a daily-consumed food group/nutrient and
energy on the original scale is defined similarly. Consider, for
example, energy, which is the $13$th dietary component and the $19$th
set of terms in the model. Let $X_{i,19,\mathrm{wkend}}$ and $X_{i,19,\mathrm{wkday}}$ be the versions of $X_{i,19,k}$ where the dummy variable has
the indicator of the weekend or weekday, respectively, and that the
recall is the first one. Then
\begin{eqnarray*}
T_{i,13,\mathrm{wkend}} &=& g_{\mathrm{tr}}^* (X_{i,19,\mathrm{wkend}}\trans\beta
_{19} + U_{i,19},\lambda_{13},\Sigma_{\epsilon,19,19} ); \\
T_{i,13,\mathrm{wkday}} &=& g_{\mathrm{tr}}^* (X_{i,19,\mathrm{wkday}}\trans\beta
_{19} + U_{i,19},\lambda_{13},\Sigma_{\epsilon,19,19} ); \\
T_{i,13} &=& (4 T_{i,13,\mathrm{wkday}} + 3 T_{i,13,\mathrm{wkend}} )/7.
\end{eqnarray*}
Similar formulae are used for the other daily-consumed foods and nutrients.

Finally, the energy-adjusted usual intakes and the HEI-2005 scores are
then obtained as in Table \ref{tab:heidescription}, using the
estimated usual intakes of the dietary components.

To find the joint distribution of usual intakes of the HEI-2005 scores,
it is convenient to use Monte Carlo methods. Recall that $w_i$ is the
sampling weight for individual $i$. Let $B$ be a large number: we set
$B = 5000$. Generate $b=1,\ldots,B$ observations $\tU_{bi} = \Normal
(0,\Sigma_u)$ and then obtain $\tT_{bi} = (T_{bi\ell})_{\ell
=1}^{13}$ by replacing $U_{ij}$ in their formulae by $U_{bij}$. With
appropriate sample weighting, the $\tT_{bi}$ can be used to estimate
joint and marginal distributions. Thus, for example, consider the total
HEI-2005 score, which is a deterministic function of the usual intakes,
say, $G(\tT_i)$. Its cumulative distribution function is estimated as
%
%e3.6 ###
\begin{equation}\label{eq:what01}
\widehat{F}(x) = \frac{\sumi\sum_{b=1}^B I\{G(\tT_{bi}) \le x\}
w_i} {\sumi\sum_{b=1}^B w_i}.
\end{equation}
Frequentist standard errors of derived quantities such as mean, median
and quantiles can be estimated using the Balanced Repeated Replication
(BRR) method [Wolter (\citeyear{Wol07})]; see Section \ref{sec5.5} for details.

%s4 ###
\section{\texorpdfstring{Comments on the approach to estimation.}{Comments on the approach to estimation}}\label{sec:howtoestimate}

Our model (\ref{eq:qkipnis04})--(\ref{eq:qkipnis05b}) is a~highly
nonlinear, mixed effects model with many latent variables and nonlinear
restrictions on the covariance matrix $\Sigma_{\epsilon}$. As seen in
Section \ref{subsec:dhei}, we can estimate relevant distributions of
usual intake in the population if we can estimate $\Sigma_u$, $\Sigma
_{\epsilon}$ and $\beta_j$ for $j=1,\ldots,19$. We have found that
working within a pseudo-likelihood Bayesian paradigm is a convenient
way to do this computation. We emphasize, however, that we are doing
this only to get frequentist parameter estimates based on the
well-known asymptotic equivalence of frequentist likelihood estimators
and Bayesian posterior means, and especially the consistency of both
[\citet{LehCas98}]. We are specifically not doing Bayesian
posterior inference, since valid Bayesian inference in a complex survey
such as NHANES is an immensely challenging task, and because
frequentist estimation and inference are the standard in the nutrition
community.

\citet{Kipetal09} were able to get estimates of parameters
separately for each food group using the nonlinear mixed effects
program NLMIXED in SAS with sampling weights. While this gives
estimates of $\beta_j$ for $j=1,\ldots,19$, it only gives us parts of the
covariance matrices $\Sigma_u$ and $\Sigma_{\epsilon}$, and not all
the entries. Using the 2001--2004 NHANES data, we have verified that our
estimates and the subset of the parameters that can be estimated by one
food group at a time using NLMIXED are in close agreement, and that
estimates of the distributions of usual intake and HEI-2005 component
scores are also in close agreement. We expect this because of the
rather large sample size in our data set. \citet{SMi} have
shown that even considering a single food group plus energy is a
challenge for the NLMIXED procedure, both in time and in convergence,
and using this method for the entire HEI-2005 constellation of dietary
components is impossible.

Full technical details of the model fitting procedure are given in
Appendices \ref{a-3}--\ref{a.6}.

Of course, our model has assumptions, for example, additivity and
homoscedasticity on a transformed scale for observed and latent
variables, normality of person-specific random effects and normality of
day-to-day variability on the transformed scale. These assumptions are
clearly not exactly correct, although our marginal model-checking
suggests to us that they are mostly not disastrously wrong. Some
reasons for this conclusion include the facts that we reproduce the
marginal distributions of the components, that comparison with  24 h
recalls shows differences that decrease when moving from one  24 h
recall to two  24 h recalls, that q-q plots of the data are fairly
satisfactory, etc. Thinking, as we do, of our work as a first step, and
not a~last step, it would be extremely interesting to make the model
more general, for example, skew-normal, skew-$t$ or Dirichlet process
distributions after transformation, and possibly directly modeling
heteroscedasticity. Such generalizations will require effort to
implement, but will speak to the robustness of the results and would be
a useful future step.

%s5 ###
\section{\texorpdfstring{Empirical work.}{Empirical work}} \label{sec:sec5}

%s5.1 ###
\subsection{\texorpdfstring{Basic analysis.}{Basic analysis}}\label{sec:sec5.1}

We analyzed data from the 2001--2004 National\break Health and Nutrition
Examination Survey (NHANES) for children ages \mbox{2--8}. The study sample
consisted of $2638$ children, among whom $1103$ children have two
 24 h recalls and the rest have only one. We used the dietary intake
data to calculate the 12 HEI-2005 components plus energy. In addition,
besides age, gender, race and interaction terms, two covariates were
employed, along with an intercept. The first was a dummy variable
indicating whether or not the recall was for a weekend day (Friday,
Saturday or Sunday) because food intakes are known to differ
systematically on weekends and weekdays. The second was a dummy
variable indicating whether the  24 h recall was the first or second
such recall, the idea being that there may be systematic differences
attributable to the repeated administration of the instrument.

%s5.2 ###
\subsection{\texorpdfstring{Contextual information.}{Contextual information}}\label{sec:sec5.2}

When we ran our program based on the variables in Table \ref
{tab:heidescription}, the results were disastrous. Mixing of the MCMC
sampler was very poor, with long sojourns in different regions.

The reason for this failure to converge depends on the context of the
dietary variables. For example, whole grains are a subset of total
grains. Thus, if someone consumes any whole grains, then necessarily,
with probability $1.0$, that person also consumes total grains. Such a
restriction cannot be handled by our model, because it would force one
of the random effects $U$ to equal infinity. A similar thing happens
for energy. Calories coming from saturated fat are a subset of total
calories, as are calories from SoFAAS, so there is a~restriction that
total calories must be greater than calories from saturated fat and
also greater than calories from SoFAAS. Since the latter sum makes up a
significant portion of calories, this restriction is not something that
our model can handle well.

Luckily, there is an easy and natural context-based solution. Instead
of using total grains in the model, we used grains that are not whole
grains, that is, refined grains, thus decoupling whole grains and total
grains, and removing the restriction mentioned above. Similarly,
instead of using total fruit, we use fruit that is not whole fruits,
that is, fruit juices. Additionally, instead of using total vegetables,
we use total vegetables excluding dark green and orange vegetables and
legumes. Finally, instead of total energy, we use total energy minus
the sum of energy from saturated fat (11\% of mean energy) and from
SoFAAS (35\% of mean energy). We recognize that there is overlap of
energy from saturated fat and energy from solid fat, but this has no
impact on our analysis since total energy has sources other than these
two. An alternative, of course, would have been to simply use total
energy minus energy from SoFAAS,

This is sufficient to estimate the distributions of interest. If, for
example, in the new data set $T_{i1}$ represents usual intake of
nonwhole fruits, and $T_{i2}$ is usual intake of whole fruits, then the
usual intake of total fruits is $T_{i1} + T_{i2}$. Similar remarks
apply for total grains and total vegetables.

With these new variables, our model mixed well and gave reasonable
looking answers that, as mentioned in Section \ref{sec:howtoestimate},
give similar results to other methods employed with smaller parts of
the data set.

%s5.3 ###
\subsection{\texorpdfstring{Estimation of the HEI-2005 scores.}{Estimation of the HEI-2005 scores}}\label{sec:5.3}

\begin{figure}

\includegraphics{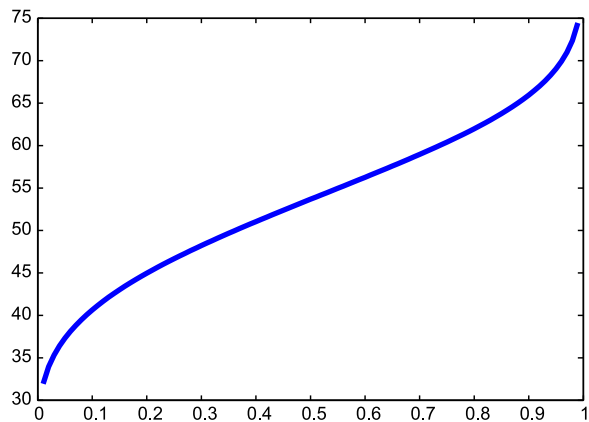}

\caption{The estimated percentiles of the HEI-2005 total score. The horizontal
axis is the percentile of interest, for example, $0.5$ refers to the
median, while the vertical axis gives percentile of the HEI-2005
scores. Standard error estimates are given in Table \protect\ref{tab:tab2}.}
\label{fig:fig1}
\end{figure}

In the \hyperref[intro]{Introduction} we posed 4 questions to which answers had not been
possible previously. The first open question concerned the distribution
of the HEI total score. Along the way toward this, Table~\ref
{tab:tab2} presents the energy-adjusted distributions of the dietary
components used in the HEI-2005. Table \ref{tab:tab3} presents the
distributions of the HEI-2005 individual component scores and the total
score, with a graphical view given in Figure \ref{fig:fig1}.
%
%t2 ###
\begin{table}
\tabcolsep=0pt
\caption{Estimated distributions of
energy-adjusted usual intakes for children aged 2--8; NHANES,~2001--2004}\label{tab:tab2}
\begin{tabular*}{\tablewidth}{@{\extracolsep{\fill}}lcd{2.2}d{2.2}d{2.2}d{2.2}d{2.2}d{2.2}d{2.2}d{2.2}@{}}
\hline
& & & \multicolumn{7}{c@{}}{\textbf{Percentile}}\\[-5pt]
&&&\multicolumn{7}{c@{}}{\hrulefill}\\
\textbf{Component} & \textbf{Units} & \multicolumn{1}{c}{\textbf{Mean}} & \multicolumn{1}{c}{\textbf{5th}} &
\multicolumn{1}{c}{\textbf{10th}} & \multicolumn{1}{c}{\textbf{25th}} & \multicolumn{1}{c}{\textbf{50th}} &
\multicolumn{1}{c}{\textbf{75th}} & \multicolumn{1}{c}{\textbf{90th}} & \multicolumn{1}{c}{\textbf{95th}} \\
\hline
Total fruit & cups/(1000 kcal)
&0.70&0.14&0.21&0.37&0.62&0.95&1.30&1.54\\
&&0.02&0.02&0.02&0.02&0.02&0.03&0.05&0.07\\[3pt]
Whole fruit & cups/(1000 kcal)
&0.31&0.04&0.07&0.14&0.26&0.42&0.61&0.73\\
&&0.02&0.01&0.01&0.02&0.02&0.03&0.04&0.06\\[3pt]
Total vegetables & cups/(1000 kcal)
&0.47&0.23&0.27&0.36&0.46&0.58&0.69&0.77\\
&&0.01&0.02&0.02&0.02&0.01&0.02&0.03&0.03\\[3pt]
DOL & cups/(1000 kcal) &0.05&0.00&0.01&0.02&0.03&0.07&0.11&0.15\\
&&0.00&0.00&0.00&0.00&0.00&0.00&0.01&0.01\\[3pt]
Total grains & ounces/(1000 kcal)
&3.32&2.35&2.54&2.87&3.28&3.72&4.16&4.45\\
&&0.05&0.08&0.07&0.06&0.05&0.06&0.08&0.10\\[3pt]
Whole grains & ounces/(1000 kcal)
&0.27&0.05&0.07&0.13&0.23&0.36&0.52&0.64\\
&&0.01&0.01&0.01&0.02&0.01&0.02&0.03&0.04\\[3pt]
Milk & cups/(1000 kcal) &0.97&0.28&0.38&0.60&0.90&1.26&1.64&1.90\\
&&0.02&0.03&0.03&0.02&0.02&0.03&0.05&0.07\\[3pt]
Meat and beans & ounces/(1000 kcal)
&1.84&1.06&1.21&1.48&1.80&2.16&2.51&2.73\\
&&0.04&0.09&0.08&0.06&0.04&0.04&0.05&0.07\\[3pt]
Oil & grams/(1000 kcal) &7.13&4.05&4.60&5.63&6.93&8.41&9.90&10.89\\
&&0.23&0.24&0.21&0.17&0.20&0.35&0.54&0.68\\[3pt]
Saturated fat & \% of energy
&11.71&8.56&9.20&10.33&11.64&13.01&14.32&15.13\\
&&0.15&0.25&0.20&0.15&0.15&0.22&0.32&0.38\\[3pt]
Sodium & grams/(1000 kcal) &1.49&1.16&1.23&1.34&1.48&1.63&1.77&1.86\\
&&0.01&0.02&0.02&0.01&0.01&0.02&0.03&0.03\\[3pt]
SoFAAS & \% of energy &36.93&27.19&29.28&32.87&36.90&40.96&44.61&46.77\\
&&0.48&0.93&0.81&0.63&0.48&0.49&0.64&0.75\\
\hline
\end{tabular*}
\legend{For each dietary component, the first line${}={}$estimate from our model,
while the second line is its BRR-estimated standard error. Here,
``DOL'' is dark green and orange vegetables and legumes. Also,
``SoFAAS'' is calories from solid fats, alcoholic beverages and added
sugars. Total Fruit, Whole Fruit, Total Vegetables, DOL and Milk are in
cups. Total Grains, Whole Grains and Meat and Beans are in ounces. Oil
and Sodium are in grams. Saturated Fat and SoFAAS are in \% of energy.
Further discussion of the size of the BRR-estimated standard errors is
given in the supplementary material [\citet{Zhaetal}].}
\end{table}

%t3 ###
\begin{table}
\caption{Estimated distributions of the usual
intake HEI-2005 scores}\label{tab:tab3}
\begin{tabular*}{\tablewidth}{@{\extracolsep{\fill}}ld{2.2}d{2.2}d{2.2}d{2.2}d{2.2}d{2.2}d{2.2}d{2.2}@{}}
\hline
& & \multicolumn{7}{c@{}}{\textbf{Percentile}}\\[-5pt]
&&\multicolumn{7}{c@{}}{\hrulefill}\\
\textbf{Component} & \multicolumn{1}{c}{\textbf{Mean}} & \multicolumn{1}{c}{\textbf{5th}} & \multicolumn{1}{c}{\textbf{10th}} &
\multicolumn{1}{c}{\textbf{25th}} & \multicolumn{1}{c}{\textbf{50th}} & \multicolumn{1}{c}{\textbf{75th}}
& \multicolumn{1}{c}{\textbf{90th}} & \multicolumn{1}{c@{}}{\textbf{95th}} \\
\hline
Total fruit &3.55&0.87&1.31&2.33&3.90&5.00&5.00&5.00\\
&0.09&0.13&0.14&0.15&0.15&0.00&0.00&0.00\\[3pt]
Whole fruit &3.14&0.49&0.82&1.71&3.24&5.00&5.00&5.00\\
&0.14&0.12&0.16&0.21&0.26&0.03&0.00&0.00\\[3pt]
Total vegetables &2.16&1.02&1.24&1.63&2.10&2.62&3.15&3.48\\
&0.06&0.10&0.10&0.07&0.06&0.07&0.12&0.16\\[3pt]
DOL &0.62&0.05&0.09&0.21&0.45&0.86&1.38&1.76\\
&0.04&0.02&0.03&0.04&0.05&0.06&0.08&0.13\\[3pt]
Total grains &4.81&3.92&4.23&4.79&5.00&5.00&5.00&5.00\\
&0.03&0.13&0.12&0.09&0.00&0.00&0.00& 0.00\\[3pt]
Whole grains &0.90&0.16&0.24&0.43&0.75&1.21&1.74&2.13\\
&0.04&0.04&0.05&0.05&0.05&0.05&0.10&0.14\\[3pt]
Milk &6.77&2.15&2.96&4.62&6.91&9.67&10.00&10.00\\
&0.12&0.23&0.22&0.18&0.17&0.25&0.00&0.00\\[3pt]
Meat and beans &7.22&4.23&4.83&5.91&7.21&8.64&10.00&10.00\\
&0.16&0.34&0.30&0.23&0.17&0.15&0.11&0.00\\[3pt]
Oil &5.92&3.37&3.83&4.69&5.77&7.01&8.25&9.07\\
&0.18&0.20&0.18&0.14&0.17&0.29&0.45&0.57\\[3pt]
Saturated fat &5.16&0.00&1.09&3.18&5.38&7.48&8.53&8.96\\
&0.21&0.35&0.51&0.35&0.24&0.23&0.13&0.16\\[3pt]
Sodium &4.52&1.25&2.05&3.31&4.62&5.83&6.85&7.44\\
&0.09&0.30&0.24&0.15&0.09&0.11&0.16&0.19\\[3pt]
SoFAAS &8.73&2.15&3.60&6.02&8.73&11.42&13.81&15.21\\
&0.32&0.50&0.42&0.33&0.32&0.42&0.54&0.62\\[6pt]
Total Score &53.50&37.42&40.74&46.73&53.68&60.36&65.87&68.96\\
&0.81&1.45&1.34&1.09&0.83&0.82&0.96&1.08\\
\hline
\end{tabular*}
\legend{For each component score, the first line${}={}$estimate from our model, while the second line is its BRR-estimated
standard error. The total score is the sum of the individual scores.
Here, ``DOL'' is dark green and orange vegetables and legumes. Also,
``SoFAAS'' is calories from solid fats, alcoholic beverages and added
sugars. Further discussion of the size of the BRR-estimated standard
errors is given in the supplementary material [\citet{Zhaetal}].}
\end{table}

%f1 ###

Table \ref{tab:tab3} presents the first estimates of the distribution
of HEI-2005 scores for a vulnerable subgroup of the population, namely,
children aged 2--8 years. A previous analysis of 2003--2004 NHANES data,
looking separately at \mbox{2--5} year olds and 6--11 year olds, was limited to
estimates of mean usual HEI-2005 scores [59.6 and 54.7, respectively;
see \citet{Funetal09}]. The mean scores noted here are comparable
to those and reinforce the notion that children's diets, on average,
are far from ideal. However, this analysis provides a more complete
picture of the state of US children's diets. By including the scores at
various percentiles, we estimate that only 5\% of children have a~score
of 69 or greater and another 10\% have scores of 41 or lower. While not
in the table, we also estimate that the 99th percentile is $74$.
This analysis suggests that virtually all children in the US have
suboptimal diets and that a sizeable fraction (10\%) have alarmingly
low scores (41 or lower.)

We have also considered whether our multivariate model fitting
procedure gives reasonable marginal answers. To check this, we note
that it is possible to use the SAS procedure NLMIXED {\it separately
for each component} to fit a~model with one episodically consumed food
group or daily consumed dietary component together with energy. The
marginal distributions of each such component done separately are quite
close to what we have reported in Table \ref{tab:tab3}, as is our
mean, which is $53.50$ compared to the mean of $53.25$ based on
analyzing one HEI-2005 component at a time with the NLMIXED procedure.
The only case where there is a mild discrepancy is in the estimated
variability of the energy-adjusted usual intake of oils, likely caused
by the NLMIXED procedure itself, which has an estimated variance $9$
times greater than our estimated variance.

Of course, it is the distribution of the HEI-2005 total score that
cannot be estimated by analysis of one component at a time.
%
%t4 ###
\begin{table}[t!]
\tabcolsep=0pt
\caption{Estimated correlation matrix for
energy-adjusted usual intakes}\label{tab:tab4}
\begin{tabular*}{\tablewidth}{@{\extracolsep{\fill}}lcd{1.2}d{1.2}d{1.2}d{2.2}d{2.2}d{2.2}d{2.2}d{2.2}d{2.2}d{2.2}d{2.2}@{}}
\hline
\textbf{Component}& \textbf{TF} & \multicolumn{1}{c}{\textbf{WF}} & \multicolumn{1}{c}{\textbf{TV}}&\multicolumn{1}{c}{\textbf{DOL}} &
\multicolumn{1}{c}{\textbf{TG}}& \multicolumn{1}{c}{\textbf{WG}}&\multicolumn{1}{c}{\textbf{Milk}}&\multicolumn{1}{c}{\textbf{Meat}}&\multicolumn{1}{c}{\textbf{Oil}}&
\multicolumn{1}{c}{\textbf{SatFat}}&\multicolumn{1}{c}{\textbf{Sodium}}&\multicolumn{1}{c@{}}{\textbf{SoFAAS}}\\
\hline
TF&1&0.76&0.07&0.41&-0.10&0.33&0.16&0.08&-0.35&-0.38&-0.25&-0.64\\
WF&&1&0.14&0.49&0.03&0.35&0.10&0.05&-0.17&-0.30&-0.20&-0.51\\
TV&&&1&0.51&-0.25&-0.23&-0.09&0.51&-0.08&0.08&0.42&-0.16\\
DOL&&&&1&-0.08&0.11&0.14&0.25&-0.06&-0.23&0.01&-0.47\\
TG &&&&&1&0.30&-0.30&-0.13&0.44&-0.36&0.17&-0.22\\
WG &&&&&&1&0.18&-0.18&-0.11&-0.29&-0.17&-0.46\\
Milk&&&&&&&1&-0.37&-0.21&0.21&-0.27&-0.21\\
Meat and beans &&&&&&&&1&-0.06&-0.08&0.39&-0.19\\
Oil&&&&&&&&&1&-0.06&0.11&0.05\\
SatFat&&&&&&&&&&1&0.09&0.46\\
Sodium&&&&&&&&&&&1&0.04\\
SoFAAS&&&&&&&&&&&&1\\
\hline
\end{tabular*}
\legend{Here TF${}={}$Total Fruits, WF${}={}$Whole
Fruits, TV${}={}$Total Vegetables, WG${}={}$Whole Grains, TG${}={}$Total Grains,
SatFat${}={}$Saturated Fat. Here, ``DOL'' is dark green and orange
vegetables and legumes. Also, ``SoFAAS'' is calories from solid fats,
alcoholic beverages and added sugars.}\vspace*{10pt}
\caption{Estimated correlations between each
individual HEI-2005 component score and the sum of the other HEI
component scores, that is, the difference of the total score and each
individual component}\label{table:scorecorr}
\begin{tabular*}{\tablewidth}{@{\extracolsep{\fill}}ld{2.2}d{2.2}d{2.2}c@{}}
\hline
& \multicolumn{1}{c}{\textbf{First  24 h}} & \multicolumn{1}{c}{\textbf{Two  24 h}} & \multicolumn{1}{c}{\textbf{Model}} & \textbf{BRR s.e.}\\
\hline
Total fruit & 0.38 & 0.44& 0.62&0.05\\
Whole fruit & 0.31 & 0.37& 0.59&0.10\\
Total vegetables & 0.09 & 0.11& 0.10&0.11\\
DOL & 0.18 & 0.24& 0.41&0.07\\
Total grains & 0.00 & 0.00& 0.06&0.11\\
Whole grains & 0.12 & 0.16& 0.53&0.08\\
Milk & -0.07 &-0.01& 0.01&0.08\\
Mean and beans & -0.03 &-0.01& -0.03&0.15\\
Oil & 0.08 & 0.05& -0.17&0.08\\
Saturated fat & 0.21 & 0.23& 0.36&0.06\\
Sodium & -0.03 & 0.05& 0.07&0.12\\
SoFAAS & 0.52 & 0.59& 0.72&0.04\\
\hline
\end{tabular*}
\legend{The column labeled ``Two 24 h'' is the naive
analysis that uses the mean of the two  24 h recalls, while the column
labeled ``First 24 h'' is the naive analysis that uses the first 24 h
recall. The column labeled ``Model'' is our analysis, and the column
labeled ``BRR s.e.'' is the estimated standard error of our estimates.
Here, ``DOL'' is dark green and orange vegetables and legumes. Also,
``SoFAAS'' is calories from solid fats, alcoholic beverages and added
sugars.}
\end{table}

There are other things that have not been computed previously that are
simple by-products of our analysis. For example, the correlations among
energy-adjusted usual intakes involving episodically consumed foods
have not been estimated previously, but this is easy for us; see Table
\ref{tab:tab4}. The estimated correlation of $-0.64$ between
energy-adjusted total fruit and energy-adjusted SoFAAS, and the $-0.47$
correlation between DOL and SoFAAS are surprisingly high.

%s5.4 ###
\subsection{\texorpdfstring{Component scores and other scores.}{Component scores and other scores}}\label{sec:5.4}

$\!\!\!$As described in the \hyperref[intro]{Introduction}, an open problem has been to estimate
the correlation between the individual score on each dietary component
and the scores of all other dietary components. In their Table 3,
Guenther et~al. (\citeyear{Gueetal08}) consider this problem, but of course they did
not have a model for usual energy adjusted intakes, and instead they
used a single  24 h recall. In Table \ref{table:scorecorr} we show the
resulting correlations using (a) a single  24 h recall; (b) the mean of
two  24 h recalls for those who have two  24 h recalls; and (c) our model
for usual intake. The numbers for the former differ from that of
Guenther et~al. (\citeyear{Gueetal08}) because we are considering here a different
population than do they. A~striking and not unexpected aspect of this
table is that for those components with nontrivial correlations, the
correlations all increase as one moves from a single  24~h recall to the
mean of two  24 h recalls and then finally to estimated usual intake.
Thus, for example, the correlation between the HEI-2005 score for total
fruit and its difference with the total score is 0.38 for a single  24 h
recall, 0.44 for the mean of two  24 h recalls and then finally 0.62 for
usual intake.
%

%t5 ###
\begin{table}[b]
\caption{Estimated distributions of
energy-adjusted usual intake for those whose total HEI-2005 total~scores~are ${\leq}50$ and ${>}50$}\label{tab:tab13}
\begin{tabular*}{\tablewidth}{@{\extracolsep{\fill}}ld{2.2}cd{2.2}d{2.2}d{2.2}d{2.2}d{2.2}d{2.2}d{2.2}@{}}
\hline
& & & \multicolumn{7}{c@{}}{\textbf{Percentile}}\\[-5pt]
&&&\multicolumn{7}{c@{}}{\hrulefill}\\
\textbf{Component} & \multicolumn{1}{c}{\textbf{Mean}} & \textbf{s.d.} & \multicolumn{1}{c}{\textbf{5th}} & \multicolumn{1}{c}{\textbf{10th}} &
\multicolumn{1}{c}{\textbf{25th}} & \multicolumn{1}{c}{\textbf{50th}} & \multicolumn{1}{c}{\textbf{75th}} & \multicolumn{1}{c}{\textbf{90th}} &
\multicolumn{1}{c@{}}{\textbf{95th}} \\
\hline
Whole fruit & & & & & & & & & \\
\quad Total score $\leq$50
&0.15&0.12&0.02&0.03&0.07&0.12&0.21&0.30&0.38\\
\quad Total score $> $50
&0.39&0.22&0.11&0.15&0.23&0.35&0.51&0.68&0.80\\[3pt]
Whole grains & & & & & & & & & \\
\quad Total score $\leq$50
&0.18&0.13&0.03&0.05&0.09&0.15&0.25&0.36&0.44\\
\quad Total score $> $50 &0.32&0.20&0.07&0.10&0.17&0.28&0.42&0.59&0.70\\[3pt]
DOL & & & & & & & & & \\
\quad Total score $\leq$50
&0.02&0.02&0.00&0.00&0.01&0.02&0.03&0.05&0.07\\
\quad  Total score $> $50
&0.06&0.05&0.01&0.01&0.03&0.05&0.09&0.13&0.17\\[3pt]
SoFAAS & & & & & & & & & \\
\quad Total score $\leq$50
&42.43&3.97&36.40&37.59&39.66&42.16&44.92&47.67&49.42\\
\quad Total score $> $50
&33.83&4.44&26.01&27.89&30.97&34.15&36.98&39.28&40.57\\[6pt]
Total Score &53.50&9.58&37.42&40.74&46.73&53.68&60.36&65.87&68.96\\
\hline
\end{tabular*}
\legend{Here, ``DOL'' is dark green and orange
vegetables and legumes. Also, ``SoFAAS'' is calories from solid fats,
alcoholic beverages and added sugars. Units of measurement are given in
Table \ref{tab:tab2}.}
\end{table}

%s5.5 ###
\subsection{\texorpdfstring{Distributions of intakes for subsets of HEI total
scores.}{Distributions of intakes for subsets of HEI total
scores}}\label{sec:5.11}

A third open question is as follows: among those whose total HEI-2005
score is $> $50 or $\leq$50, what is the distribution of
energy-adjusted usual intake of whole grains, whole fruits, dark green
and orange vegetables and legumes (DOL) and calories from solid fats,
alcoholic beverages and added sugars (SoFAAS)? This follows naturally
from our method. Following (\ref{eq:what01}), let $G_1(\tT_{bi})$ be
energy adjusted usual intake and let $G_2(\tT_{bi})$ be the HEI total
score. Then the distributions in question for when the total HEI-2005
score is $> $50 can be estimated as $\widehat{F}(x) = \sumi\sum
_{b=1}^B w_i I\{G_1(\tT_{bi}) \le x\}I\{G_2(\tT_{bi}) > 50\} / \sumi
\sum_{b=1}^B w_i I\{G_2(\tT_{bi}) > 50\}$.

\begin{figure}

\includegraphics{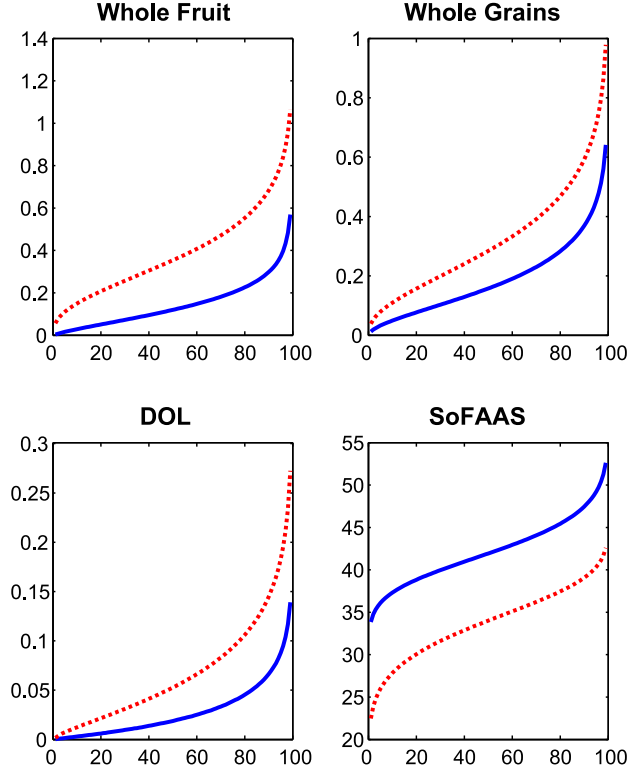}

\caption{The estimated percentiles of the energy-adjusted usual intakes for
Whole fruits \textup{(top left)} in cups/(1,000 kcal), Whole grains \textup{(top right)}
in ounces/(1,000 kcal), DOL \textup{(bottom left)} in cups/ (1,000~kcal) and
calories from SoFAAS \textup{(bottom right)} in \% of Energy. The solid lines
are for those whose usual HEI-2005 total score is $\leq$50, that is,
poorer diets, while the dashed lines are for those whose usual HEI-2005
total score is $>$50, that is, better diets.}\label{fig:fig2}
\end{figure}

The results are provided in Table \ref{tab:tab13}, with a graphical
view in Figure \ref{fig:fig2}. The results show that those who have
poorer diets with usual HEI-2005 total score $\leq50$ are consistently
eating poorer diets, that is, less whole fruits, less whole grains and
less DOL, but higher SoFAAS.

%f2 ###

%s5.6 ###
\subsection{\texorpdfstring{Dietary consistency.}{Dietary consistency}}\label{sec:5.12}

\begin{figure}

\includegraphics{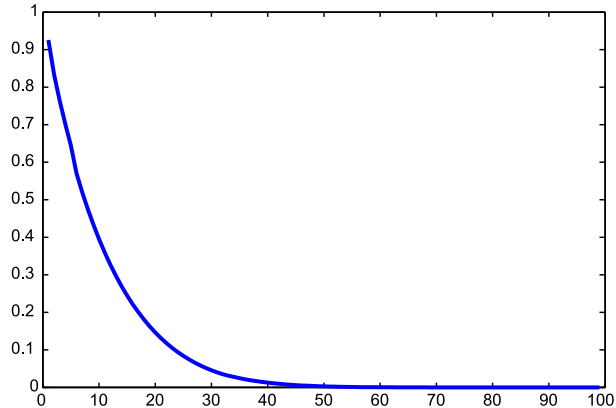}

\caption{The Y-axis gives the estimated probabilities of exceeding the $\kappa$
(X-axis) percentile on all 12 HEI components, for $\kappa=
1,2,\ldots,99$; see Section \protect\ref{sec:5.12}.}\label{fig:fig3}
\end{figure}

We stated in the \hyperref[intro]{Introduction} that it is interesting to understand the
percentage of children whose usual intake HEI score exceeds the median
HEI score on all 12 HEI components. Those median scores, say, $(\kappa
_1,\ldots,\kappa_{12})$, are estimated in Table \ref{tab:tab3}. If
$G_j(\tT_{bi})$ is the HEI component score for episodically consumed
food $j$, then following (\ref{eq:what01}) the quantity in question
can be estimated as $\sumi\sum_{b=1}^B w_i\prod_{j=1}^6 I\{G_j(\tT
_{bi}) \geq\kappa_j\} / \sumi\sum_{b=1}^B w_i$. We estimate that
the percentage is 6\%, woefully small. The percentage of children whose
usual intake HEI score exceeds the median HEI score on all 12 HEI
components is 0.24\%. Figure \ref{fig:fig3} gives the estimated
probabilities of exceeding the $\kappa$ percentile on all 12 HEI
components simultaneously, for $\kappa= 1,2,\ldots,99$.

%f3 ###

%s5.7 ###
\subsection{\texorpdfstring{Uncertainty quantification.}{Uncertainty quantification}}\label{sec5.5}
The BRR standard errors of HEI-2005 components' adjusted usual intakes
and scores are shown in Tables \ref{tab:tab2} and~\ref{tab:tab3}. The
BRR weights are only used in variance calculations. Once we have
estimated some quantity, say, $\widehat{\theta}$, from the sample
using sample weight, we will need to compute the same quantity using,
in succession, the 32 BRR weights. This will give us 32 estimates
$\widehat{\theta}_1, \widehat{\theta}_2, \ldots,\widehat{\theta
}_{32}$. The BRR estimate for the variance of $\widehat{\theta}$ is
$(32\times0.49)^{-1} \sum_{p=1}^{32}(\widehat{\theta}_p -
\widehat{\theta})^2$. The 32 in the denominator is for the 32
different estimates from the 32 different sets of weights, and the
$0.49$ is the square of the perturbation factor used to construct the
BRR weight sets [Wolter (\citeyear{Wol07})].

%s6 ###
\section{\texorpdfstring{Further discussion of the analysis.}{Further discussion of the analysis}}\label{sec:sec6}

%s6.1 ###
\subsection{\texorpdfstring{Never consumers.}{Never consumers}}\label{sec6.1}

An aspect of the modeling that we have not discussed is the possibility
that some people never, ever consume an episodically consumed dietary
component. Our model does not allow for this, for general reasons and
for reasons that are specific to our data analysis.

It is in principle possible to add an additional modeling step for
nonconsumers, via fixed effects probit regression, but we do not think
this is a~practical issue in our case, for two reasons:
\begin{itemize}
\item The first is that the HEI-2005 is based on 6 episodically
consumed dietary components, namely, total fruit, whole fruit, whole
grains, total vegetables, DOL and milk, the latter of which includes
cheese, yogurt and soy beverages. None of these are ``lifestyle
adverse,'' unlike, say, alcohol. While 40\% of the responses for whole
fruits, for example, equal zero, the percentage of children who never
eat any whole fruits at all is likely to be minuscule.
\item Even if one disputes whether there are very few individuals who
never consume one of the dietary components, then it necessarily
follows that we have \textit{overestimated} the HEI-2005 total
scores, and, hence, the estimates of the proportion of individuals with
alarmingly low HEI scores are \textit{deflated}, and not inflated.
The reason is that our model suggests everyone has a~positive usual
intake of the 6 episodically consumed dietary components. Since the
HEI-2005 score components are nondecreasing functions of usual intake
of the episodically consumed dietary components, this would mean that
we overestimate the HEI-2005 total score.
\end{itemize}

\subsection{\texorpdfstring{Computing and data.}{Computing and data}}\label{canddata}

Our programs were written in Matlab. The programs, along with the
NHANES data we used, are available in the \textit{Annals of Applied
Statistics} online archive.
Although a much smaller amount of computing effort yields similar
results, using 70,000 MCMC steps with a~burn-in of 20,000 takes
approximately 10 hours on a Linux server.

We also estimated the Monte Carlo standard error which is defined by
\citet{FleHarJon08} as $\widehat\sigma_g / \sqrt{n}$, where $n$ is
the total of iterations, and $n=ab$, where $a$ is the number of blocks
and $b$ is the block size, and where\vspace*{-4pt}
\[
\bar{Y_j} = b^{-1} \sum_{i={(j-1)b+1}}^{jb} g(X_i) \qquad \mbox{for }j= 1, \ldots, a.
\]
The batch means estimate of $\sigma_g^2$ is\vspace*{-1pt}
\[
\widehat\sigma_g^2 = \frac{b}{a-1}\sum_{j=1}^a(\bar Y_j - \bar g_n)^2.
\vspace*{-1pt}
\]
The ratio of the Monte Carlo standard error to the estimated standard
deviation of the estimated parameters averages $3.4\%$ for $\Sigma_u$
and $1.7\%$ for $\beta$.

Because of the public health importance of the problem, the National
Cancer Institute has contracted for the creation of a SAS program that
performs our analysis. It will allow any number of episodically and
daily consumed dietary components. The first draft of this program,
written independently in a different programming language, gives almost
identical results to what we have obtained, at least suggesting that
our results are not the product of a programming error.

\vspace*{-1pt}
%s7 ###
\section{\texorpdfstring{Discussion.}{Discussion}}\label{sec:sec7}
\vspace*{-1pt}

%s7.1 ###
\subsection{\texorpdfstring{Transformations.}{Transformations}}\label{sec:trans}

In Appendix \ref{esttransformation} we describe how we estimated the
transformation parameters as a separate component-wise calculation. We
have done some analyses where we simultaneously transform each
component, and found very little difference with our results. However,
the comp-\vfill\eject\noindent uting time to implement this is extremely high, because of the
fact that different transformations make data on different scales, so
we have to compute the usual intakes at each step in the MCMC, and not
just at the end.

%s7.2 ###
\subsection{\texorpdfstring{What have we learned that is new.}{What have we learned that is new}}\label{discuss1}
There are many important questions in dietary assessment that have not
been able to be answered because of a lack of multivariate models for
complex, zero-inflated data with measurement errors and a lack of
ability to fit such multivariate models. Nutrients and foods are not
consumed in isolation, but rather as part of a broader pattern of
eating. There is reason to believe that these various dietary
components interact with one another in their effect on health,
sometimes working synergistically and sometimes in opposition.
Nonetheless, simply characterizing various patterns of eating has
presented enormous statistical challenge. Until now, descriptive
statistics on the HEI-2005 have been limited to examination of either
the total scores or only a single energy-adjusted component at a time.
This has precluded characterization of various patterns of dietary
quality as well as any subsequent analyses of how such patterns might
relate to health.

This methodology presented in this paper presents a workable solution
to these problems which has already proven valuable. In May 2010, just
as we were submitting the paper, a White House Task Force on Childhood
Obesity created a report. They had wanted to set a goal of all children
having a total HEI score of 80 or more by 2030, but when they learned
we estimated only 10\% of the children ages 2--8 had a score of 66 or
higher, they decided to set a more realistic target. The facility to
estimate distributions of the multiple component scores simultaneously
will be important in tracking progress toward that goal.

%s7.3 ###
\subsection{In what other arenas will our work have impact?}\label{discuss2}
There are many other important problems where multivariate models such
as ours will be important. One such problem arises when studying the
relationship between multiple dietary components or dietary patterns
and health outcomes. Traditionally, for cost reasons, large cohort
studies have used a food frequency questionnaire (FFQ) to measure
dietary intake, sometimes with a small calibration study including
short-term measures such as  24 h recalls. However, there is a new
web-based instrument called the Automated Self-administered 24-hour
Dietary Recall (ASA24$^{\mathrm{TM}}$) (see
\href{http://riskfactor.cancer.gov/tools/instruments/asa24}{http://riskfactor.cancer.gov/}\break
\href{http://riskfactor.cancer.gov/tools/instruments/asa24}{tools/instruments/asa24}), which has been
proposed to replace or at least supplement the FFQ and which is
currently undergoing extensive testing. The dietary data we will see
then is what we have called $Y_{ijk}$, that is,  24~h recall data. In
order to correct relative risk estimates for the measurement error
inherent in the ASA24$^{\mathrm{TM}}$, regression calibration [\citet{Caretal06}] will almost certainly be the method of choice, as it is in
most of nutritional epidemiology. This method attempts to produce an
estimate of the regression of usual intake on the observed intakes, and
then to use these estimates in Cox and logistic regression for the
health outcome. In order to perform this regression, a multivariate
measurement error model will be required, since the regression is on
\textit{all} the observed dietary intake components in the
regression model measured by the ASA24$^{\mathrm{TM}}$, and not on each
individual component. Our methodology is easily extended to address
this problem.

\begin{appendix}
\section*{\texorpdfstring{Appendix: Details of the fitting procedure}{Appendix: Details of the fitting procedure}}\label{app}
\setcounter{equation}{0}

In this \hyperref[app]{Appendix} we give the full details of the model fitting procedure.

%s7.4 ###
\subsection{\texorpdfstring{Notational convention.}{Notational convention}}\label{a-3}

In our example, age was standardized to have mean $0.0$ and variance
$1.0$, to improve numerical stability.

As described in Section \ref{sec:sec2_1}, the observed, transformed
nonzero  24 h recalls were standardized to have mean $0.0$ and variance
$2.0$. More precisely, for $\ell= 1, 2, \ldots, 6$, we first transformed
the nonzero food group data as $Z_{i,2\ell,k} = g(Y_{i,2\ell
,k},\lambda_{\ell})$, and then we standardized these data as
$Q_{i,2\ell,k} = \sqrt{2}\{Z_{i,2\ell,k} - \mu(\lambda_{\ell})\}
/\sigma(\lambda_{\ell})$, where $\{\mu(\lambda_{\ell}),\sigma
(\lambda_{\ell})\}$ are the mean and standard deviation of the
nonzero food intakes $Z_{i,2\ell,k}$. Similarly, for nonepisodically
consumed dietary components and energy we transformed to $Z_{i,6+\ell
,k} = g(Y_{i,6+\ell,k},\lambda_{\ell})$ for $\ell=7,\ldots,13$, and
then standardized to $Q_{i,6+\ell,k} = \sqrt{2}\{Z_{i,6+\ell,k} -
\mu(\lambda_{\ell})\}/\sigma(\lambda_{\ell})$. Of course, whether
the food group is consumed or not is $Q_{i,2\ell-1,k} = Y_{i,2\ell
-1,k}$ for $\ell= 1,\ldots,6$. Collected, the data are $\tQ_{ik} =
(Q_{ijk})_{j=1}^{19}$. The terms $\{\mu(\lambda_{\ell}),\break\sigma
(\lambda_{\ell})\}$ are not random variables but are merely constants
used for standardization, and we need not consider inference for them.
Back-transformation is discussed in Appendix \ref{backtransformation}.

%s7.5 ###
\subsection{\texorpdfstring{Prior distributions.}{Prior distributions}}\label{a-2}
Because the data were standardized, we used the following conventions:
\begin{itemize}
\item The prior for all $\beta_j$ were normal with mean zero and
variance $100$.
\item The prior for $\Sigma_{u}$ was exchangeable with diagonal
entries all equal to $1.0$ and correlations all equal to $0.50$. There
were 21 degrees of freedom in the inverse Wishart prior, that is,
$m_{u} = 21$. Thus, the prior is $\IW\{(m_{u}-19-1)\Sigma_{u,\mathrm{prior}}, m_{u}\}$. We experimented with this prior by using zero
correlation, and the results were essentially unchanged.
\item The prior for $r_k$ is $\operatorname{Uniform}[-1, 1]$. Set the initial value:
$r_k = 0$, $k = 1,\ldots, 5$.
\item The prior for $\theta_{k}$ is $\operatorname{Uniform}[-\pi, \pi]$. Set the
initial value: $\theta_{k} = 0$, $k = 1,\ldots, 25$.
\item The priors for $v_{22},v_{44}, \ldots, v_{12,12}$ and
$v_{13,13}, \ldots, v_{19,19}$ were $\operatorname{Uniform}[-3,3]$. Set the initial
values: $v_{22} = v_{44} = \cdots= v_{12,12}=v_{13,13}= \cdots=
v_{19,19} =1$.
\item For the rest of the nondiagonal $v_{ij}$'s which could not be
determined by the restrictions, we used $\operatorname{Uniform}[-3,3]$ priors. Set the
initial values to be~0.
\end{itemize}

The constraints on $\Sigma_{\epsilon}$ are nonlinear, and our
parameterization enforces them easily without having to have prior
distributions for the original parameterization that satisfy the
nonlinear constraints.

The key thing that makes things work well with the other components of
the matrix $V$ with $\Sigma_{\epsilon} = V V\trans$ is that we have
standardized the data as described in Appendix \ref{a-3}. With this
standardization, things become much nicer. For example, the variance of
the $\epsilon$'s for energy is $\sum_{j=1}^{19}v_{19,j}^2$. However,
since the sample variance for energy is standardized to equal $2.0$, we
simply just need to make priors for $v_{19,j}$ be uniform on a modest
range to have real flexibility.

%s7.6 ###
\subsection{\texorpdfstring{Generating starting values for the latent variables.}{Generating starting values for the latent variables}}\label{a-1}

While we observe $\tQ_{ik}$, in the MCMC we need to generate starting
values for the latent variables $\tW_{ik} = (W_{ijk})_{j=1}^{19}$ to
initiate the MCMC:
\begin{itemize}
\item For nutrients and energy, $Q_{ijk} = W_{ijk}$, no data need be
generated, $j=13,\ldots, 19$.
\item For the amounts, $Q_{i2k}$, $Q_{i4k}$, $Q_{i6k}$, $Q_{i8k}$,
$Q_{i,10,k}$ and $Q_{i,12,k}$, we set $W_{i2k} = Q_{i2k}$, $W_{i4k} =
Q_{i4k}$, $W_{i6k} = Q_{i6k}$, $W_{i8k} = Q_{i8k}$, $W_{i,10,k} =
Q_{i,10,k}$ and $W_{i,12,k} = Q_{i,12,k}$.
\item For consumption, we generate $\tU_i$ as normally distributed
with mean zero and covariance matrix given as the prior covariance
matrix for $\Sigma_u$. For $\ell= 1, \ldots, 6$, we also compute
$z_{ik} = |X_{i, 2\ell-1,k}\trans\beta_{2\ell-1,\mathrm{prior}} + U_{i,
2\ell-1} + \mathcal{Z}_{ik}|$, where $\mathcal{Z}_{ik} = \Normal(0,1)$ are
generated independently. We then set $W_{i, 2\ell-1, k} = z_{ik}Q_{i,
2\ell-1, k} - z_{ik}(1-Q_{i, 2\ell-1, k})$.
\item Finally, we then updated $\tW_{ik}$ by a single application of
the updates given in Appendix \ref{subsec:wi1}.
\end{itemize}

%s7.7 ###
\subsection{\texorpdfstring{Complete data loglikelihood.}{Complete data loglikelihood}}\label{sec4.5}
Let $J=19$. The complete data include the indicators of whether a food
was consumed, the $W$ variables and the random effect $U$ variables.
The loglikelihood of the complete data is
\begin{eqnarray*}
&& \sumell\sumi\sumk\log\{Q_{i, 2\ell-1, k}I(W_{i, 2\ell-1,k} >
0) \\
&&\hphantom{\sumell\sumi\sumk\log\{}{} + (1-Q_{i, 2\ell-1, k})I(W_{i, 2\ell-1, k}<0)\}\\
&&\qquad {} + \Biggl(\sumi w_i /2\Biggr)\log(|\Sigma_u^{-1}|)-(1/2)\sumi w_i \tU_i\trans
\Sigma_u^{-1}\tU_i \\
&&\qquad {} -(1/2) \sumjJ(\beta_j - \beta_{j,\mathrm{prior}})\trans\Omega_{\beta
,j}^{-1}(\beta_j - \beta_{j,\mathrm{prior}}) \\
&&\qquad {} + \{(m_{u}+J+1)/2\}\log(|\Sigma_{u}^{-1}|) - \{(m_{u}-J-1)/2\}
\trace(\Sigma_{u,\mathrm{prior}}\Sigma_{u}^{-1}) \\
&&\qquad {} - (1/2)\sumi w_i m_i \log\{(v_{22}^2 v_{44}^2 v_{66}^2 v_{88}^2
v_{10,10}^2 v_{12,12}^2 v_{13,13}^2 \cdots v_{JJ}^2)\prod
_{q=1}^{5}(1-r_q^2)\}\\
&&\qquad {} -(1/2) \sumi w_i \sumk\{\tW_{ik} - (X_{i1k}\trans\beta
_1,\ldots,X_{iJk}\trans\beta_{J})\trans- \tU_i\}\trans\Sigma
_{\epsilon}^{-1} \\
&&\hphantom{\qquad {} -(1/2) \sumi w_i \sumk}
{}\times\{\tW_{ik} - (X_{i1k}\trans\beta_1,\ldots,X_{iJk}\trans\beta
_{J})\trans- \tU_i\}.
\end{eqnarray*}
We used Gibbs sampling to update this complete data loglikelihood, the
details for which are given in subsequent appendices. The weights $w_i$
are integers and are used here in a pseudo-likelihood fashion. One can
also think of this as expanding each individual into $w_i$ individuals,
each with the same observed data but different latent variables. For
computational convenience, since we are only asking for a frequentist
estimator and not doing full Bayesian inference, the latent variables
in the process are generated once for each individual. Estimates of
$\Sigma_u$, $\Sigma_{\epsilon}$ and $\beta_j$ for $j=1,\ldots,J$ were
computed as the means from the Gibbs samples. Once again, we emphasize
that we are not doing a proper Bayesian analysis, but only using MCMC
techniques to obtain a frequentist estimate, with uncertainty assessed
using the frequentist BRR method.

%s7.8 ###
\subsection{\texorpdfstring{Complete conditionals for $r_q$, $\theta_{q}$ and
$v_{pq}$.}{Complete conditionals for $r_q$, $\theta_{q}$ and
$v_{pq}$}}\label{a.1}

Except for irrelevant constants, the complete conditional for $r_q$ ($q
= 1, \ldots, 5$) is
\begin{eqnarray*}
\log[r_q|\rest]
&=& - \frac{1}{2}\sumi w_i m_i \log(1-r_q^2) \\
&&{}-\frac{1}{2} \sumi w_i \sumk\{\tW_{ik}
- (X_{i1k}\trans\beta_1,\ldots,X_{i,19,k}\trans\beta_{19})\trans- \tU
_i\}\trans\\
&& \hphantom{{}-\frac{1}{2} \sumi w_i \sumk}{} \times\Sigma_{\epsilon}^{-1} \{\tW_{ik} -
(X_{i1k}\trans\beta_1,\ldots,X_{i,19,k}\trans\beta_{19})\trans- \tU
_i\}.
\end{eqnarray*}

Except for irrelevant constants, the complete conditionals for $v_{qq}$
($q= 2, 4, 6, 8, 10, 12, 13, \ldots, 19$) are
\begin{eqnarray*}
\log[v_{qq}|\rest]
&=& - \frac{1}{2}\sumi w_i m_i \log(v_{qq}^2) \\
&&{}-\frac{1}{2} \sumi w_i \sumk\{\tW_{ik}
- (X_{i1k}\trans\beta_1,\ldots,X_{i,19,k}\trans\beta_{19})\trans- \tU
_i\}\trans\\
&& \hphantom{{}-\frac{1}{2} \sumi w_i \sumk}{} \times\Sigma_{\epsilon}^{-1}\{\tW_{ik} -
(X_{i1k}\trans\beta_1,\ldots,X_{i,19,k}\trans\beta_{19})\trans- \tU
_i\}.
\end{eqnarray*}

Except for irrelevant constants, the compete conditionals for $\theta
_{q}$ ($q = 1, \ldots, 25$) and nondiagonal free parameters $v_{pq}$ are
\begin{eqnarray*}
\log[x|\rest] &=& -\frac{1}{2} \sumi w_i \sumk\{\tW_{ik}
- (X_{i1k}\trans\beta_1,\ldots,X_{i,19,k}\trans\beta_{19})\trans- \tU
_i\}\trans\\
&& \hskip25mm \times\Sigma_{\epsilon}^{-1}\{\tW_{ik} -
(X_{i1k}\trans\beta_1,\ldots,X_{i,19,k}\trans\beta_{19})\trans- \tU
_i\}.
\end{eqnarray*}
The full conditionals do not have an explicit form, so we use a
Metropolis--Hastings within a Gibbs sampler to generate it:

\begin{itemize}
\item$r_q$ ($q = 1, \ldots, 5$).
We discretize the values of $r_q$ to the set $\{-0.99 + 2 \times0.99
(j-1)/(M-1)\}$, where $j=1,\ldots,M$ and we choose $M = 41$.

Proposal: The current value is $r_{q,t}$. The proposed value of
$r_{q,t+1}$ is selected randomly from the current value and the two
nearest neighbors of $r_{q,t}$. Then $r_{q,t+1}$ is accepted with
probability $\min\{ 1,g(r_{q,t+1})/g(r_{q,t})\}$, where
\begin{eqnarray*}
g(y)
&\propto&(1-y^2)^{-{1}/{2}\sumi w_i m_i}\\
&&{} \!\!\!\times\exp\Biggl[ \!- \frac{1}{2} \sumi w_i \!\sumk\{\tW_{ik}
\!-\! (X_{i1k}\trans\beta_1,\ldots,X_{i,19,k}\trans\beta_{19})\trans\!-\! \tU
_i\}\trans
\Sigma_{\epsilon}^{-1}(\bullet) \Biggr]\!,
\end{eqnarray*}
where here and in what follows, for any $A$, $A\trans\Sigma_{\epsilon
}^{-1} (\bullet) = A\trans\Sigma_{\epsilon}^{-1} A$.

\item$\theta_{q}$ ($q = 1, \ldots, 25$).
We discretize similarly as above.

Proposal: The current value is $\theta_{q,t}$. The proposed value
$\theta_{q,t+1}$ is selected randomly from the current value and the
two nearest neighbors of $\theta_{q,t}$. Then $\theta_{q,t+1}$ is
accepted with probability $\min\{ 1,g(\theta_{q,t+1})/g(\theta
_{q,t})\}$,
where
\[
g(y)\propto \exp\Biggl[ - \frac{1}{2} \sumi w_i \sumk\{\tW_{ik}
- (X_{i1k}\trans\beta_1,\ldots,X_{i,19,k}\trans\beta_{19})\trans- \tU
_i\}\trans
\Sigma_{\epsilon}^{-1}(\bullet) \Biggr].
\]

\item$v_{qq}$ ($q= 2, 4, 6, 8, 10, 12, 13, \ldots, 19$).
Proposal: The current value is $v_{qq,t}$. A~candidate $v_{qq,t+1}$ is
generated from the Uniform distribution of length 0.4 with mean $v_{qq, t}$.
The candidate value $v_{qq,t+1}$ is accepted with probability $\min\{
1, g(v_{qq,t+1})/g(v_{qq, t})\}$, where
\begin{eqnarray*}
g(y)& \propto& y^{-\sumi w_i m_i} \\
&& {}\!\!\!\times\exp\Biggl[ \!- \frac{1}{2} \sumi w_i \!\sumk\{\tW_{ik}
\!-\! (X_{i1k}\trans\beta_1,\ldots,X_{i,19,k}\trans\beta_{19})\trans\!-\! \tU
_i\}\trans
\Sigma_{\epsilon}^{-1}(\bullet) \Biggr]\!.
\end{eqnarray*}

\item Nondiagonal free parameters $v_{pq}$.
Proposal: The current value is $v_{pq,t}$. The candidate value
$v_{pq,t+1}$ is generated from the Uniform distribution of length 0.4
with mean $v_{pq,t}$. The candidate value is accepted with probability
$\min\{1, g(v_{pq,t+1})/g(v_{pq,t})\}$, where
\[
g(y)\propto
\exp\Biggl[ - \frac{1}{2} \sumi w_i \sumk\{\tW_{ik}- (X_{i1k}\trans
\beta_1,\ldots,X_{i,19,k}\trans\beta_{19})\trans- \tU_i\}\trans
\Sigma_{\epsilon}^{-1}(\bullet) \Biggr].
\]
\end{itemize}

%s7.9 ###
\subsection{\texorpdfstring{Complete conditionals for $\Sigma_u$.}{Complete conditionals for $\Sigma_u$}}\label{subsec:sigmau}
The dimension of the covariance matrices is $J = 19$. By inspection,
the complete conditional for $\Sigma_u$ is
\[
[\Sigma_u|\rest] = \IW\Biggl\{(m_{u}-J-1)\Sigma_{u,\mathrm{prior}} + \sumi
w_i \tU_i\tU_i\trans,n+m_u\Biggr\},
\]
where here $\IW
= \mathrm{the}$ Inverse-Wishart distribution. The density of $\IW(\Omega,m)$
for a $J\times J$ random variable is
\[
\IW(\Omega,m) = f(Q|\Omega,m)
\propto|Q|^{-(m+J+1)/2}\exp\bigl\{-\tfrac{1}{2}\trace(\Omega Q^{-1})\bigr\}.
\]
This has expectation $\Omega/(m-J-1)$.

%s7.10 ###
\subsection{\texorpdfstring{Complete conditionals for $\beta$.}{Complete conditionals for $\beta$}}\label{subsec:beta}
Let the elements of $\Sigma_{\epsilon}^{-1}$ be
$\sigma_{\epsilon}^{j\ell}$. For any~$j$, except for irrelevant
constants,
\begin{eqnarray*}\log[\beta_j | \rest]
&=& -\frac{1}{2}(\beta_j - \beta_{j,\mathrm{prior}})\trans\Omega_{\beta
,j}^{-1}(\beta_j - \beta_{j,\mathrm{prior}}) \\
&&{} -\frac{1}{2} \sumi w_i \sumk(W_{ijk} - X_{ijk}\trans\beta_j -
U_{ij})^2\sigma_{\epsilon}^{jj} \\
&&{} - \sumi w_i \sumk \sum_{\ell\neq j} \sigma_{\epsilon
}^{j\ell}
(W_{ijk} - X_{ijk}\trans\beta_j - U_{ij})  (W_{i\ell k} - X_{i\ell k}\trans\beta_\ell-
U_{i\ell}) \\
&=& \mathcal{C}_1\trans\beta_j - \frac{1}{2}\beta_j\trans\mathcal{C}_2^{-1}\beta_j,
\end{eqnarray*}
which implies $ [\beta_j | \rest] =
\Normal(\mathcal{C}_2 \mathcal{C}_1,\mathcal{C}_2)$, where
\begin{eqnarray*}
\mathcal{C}_2&=& \Biggl(\Omega_{\beta,j}^{-1} + \sumi w_i \sigma_{\epsilon
}^{jj}\sumk X_{ijk}X_{ijk}\trans\Biggr)^{-1}; \\
\mathcal{C}_1&=& \Omega_{\beta,j}^{-1}\beta_{j,\mathrm{prior}} + \sumi w_i
\sumk\sigma_{\epsilon}^{jj}X_{ijk}(W_{ijk}-U_{ij}) \\
&&{} + \sumi w_i \sumk \sum_{\ell\neq j}
\sigma_{\epsilon}^{j\ell}(W_{i\ell k} - X_{i\ell k}\trans\beta
_\ell-
U_{i\ell}) X_{ijk}.
\end{eqnarray*}

%s7.11 ###
\subsection{\texorpdfstring{Complete conditionals for $\tU_{i}$.}{Complete conditionals for $\tU_{i}$}}\label{subsec:ui}

The NHANES 2001--2004 weights are integers, representing the number of
children that each sampled child represents. Thus, as described
therein, the loglikelihood in Section \ref{sec4.5} could also be
rewritten equivalently by developing $w_i$ pseudo-children, each with
the same observed data values. It thus does not make sense to use the
weights to generate an individual $\tU_{i}$. Instead, as described in
Section \ref{sec4.5}, for computational convenience for generating a
$\tU_{i}$ to represent $w_i$ children, we set the weight for that
child temporarily $= 1.0$. Then, except for irrelevant
constants,
\begin{eqnarray*}
\log[\tU_{i}|\rest]&=& -\frac{1}{2} w_i\tU_i\trans\Sigma
_u^{-1}\tU_i \\[-2pt]
&& {}-\frac{1}{2}w_i\sumk\{\tW_{ik} -
(X_{i1k}\trans\beta_1,\ldots,X_{i,19,k}\trans\beta_{19})\trans-
\tU_i\}\trans\Sigma_{\epsilon}^{-1} \\[-2pt]
&&\hphantom{{}-\frac{1}{2}w_i\sumk}
{}\times\{\tW_{ik} - (X_{i1k}\trans\beta_1,\ldots,X_{i,19,k}\trans\beta
_{19})\trans- \tU_i\} \\[-2pt]
&=& \mathcal{C}_1\trans\tU_i - \frac{1}{2}\tU_i\trans\mathcal{C}_2^{-1}\tU_i.
\end{eqnarray*}
Remembering that for purposes of this section we are setting $w_i =
1.0$, this implies that $[\tU_{i}|\rest] = \Normal(\mathcal{C}_2
\mathcal{C}_1,\mathcal{C}_2)$, where
\begin{eqnarray*}
\mathcal{C}_2&=& (\Sigma_u^{-1} + m_i\Sigma_{\epsilon}^{-1})^{-1}; \\
\mathcal{C}_1&=& \sumk\Sigma_{\epsilon}^{-1} \{\tW_{ik} -
(X_{i1k}\trans\beta_1,\ldots,X_{i,19,k}\trans\beta_{19})\trans\}.
\end{eqnarray*}

%s7.12 ###
\subsection{\texorpdfstring{Complete conditional for $W_{i\ell k}$, $\ell=
1,3,5,7,9,11$.}{Complete conditional for $W_{i\ell k}$, $\ell=
1,3,5,7,9,11$}}\label{subsec:wi1}

Here we do the complete conditional for $W_{i\ell k}$ with $\ell=
1,3,5,7,9,11$. Except for irrelevant constants,
\begin{eqnarray*}
\log[W_{i\ell k}|\rest] &=& \log\{Q_{i\ell k}I(W_{i\ell k} > 0) +
(1-Q_{i\ell k})I(W_{i\ell k}<0)\} \\
&&{} -\frac{1}{2}w_i
(W_{i1 k}-X_{i1k }\trans\beta_1
-U_{i1},\ldots,W_{i,19,k}-X_{i,19,k}\trans\beta_{19}- U_{i,19})\\
&&\hphantom{{}-}{}\times\Sigma^{-1}_{\epsilon} (\bullet)\trans\\
&=& \log\{Q_{i\ell k}I(W_{i\ell k} > 0) + (1-Q_{i\ell k})I(W_{i\ell
k}<0)\} \\
&&{} -\frac{1}{2}w_i \sigma_{\epsilon}^{\ell\ell} (W_{i\ell
k}-X_{i\ell k}\trans\beta_\ell-U_{i\ell})^2 \\
&& {}- w_i \sum_{j\neq\ell} \sigma_{\epsilon}^{\ell j}
(W_{i\ell k}-X_{i\ell k}\trans\beta_\ell-U_{i\ell})
(W_{ijk}-X_{ijk}\trans\beta_j-U_{ij}) \\
&=& \log\{Q_{i\ell k}I(W_{i\ell k} > 0) + (1-Q_{i\ell k})I(W_{i\ell
k}<0)\}
+ \mathcal{C}_1 W_{i\ell k} \\
&&{} - \frac{1}{2} W_{i\ell k}^2 \mathcal{C}_2^{-1},
\end{eqnarray*}
where, using the convention of Appendix \ref{subsec:ui},
\begin{eqnarray*}
\mathcal{C}_2 &=& 1/(\sigma_{\epsilon}^{\ell\ell}) \\
\mathcal{C}_1 &=& \sigma_{\epsilon}^{\ell\ell} (X_{i\ell k}\trans
\beta_\ell+U_{i\ell})
- \sum_{j\neq\ell} \sigma_{\epsilon}^{\ell
j}(W_{ijk}-X_{ijk}\trans\beta_j-U_{ij}).
\end{eqnarray*}
If we use the notation $\mathrm{TN}_+(\mu,\sigma,c)$ for a normal
random variable with mean~$\mu$ and standard deviation $\sigma$ that
is truncated from the left at $c$, and similarly use $\mathrm{TN}_-(\mu
,\sigma,c)$ when truncation is from the right at $c$, then it follows
that with $\mu= \mathcal{C}_2\mathcal{C}_1$ and $\sigma= \mathcal{C}_2^{1/2}$,
\begin{eqnarray*}
[W_{i\ell k}|\rest] &=& Q_{i\ell k} \mathrm{TN}_+(\mu,\sigma,0) +
(1-Q_{i\ell k})\mathrm{TN}_-(\mu,\sigma,0) \\
&=& \mu+ Q_{i\ell k} \mathrm{TN}_+(0,\sigma,-\mu) + (1-Q_{i\ell
k})\mathrm{TN}_-(0,\sigma,-\mu) \\
&=& \mu+ Q_{i\ell k} \mathrm{TN}_+(0,\sigma,-\mu) - (1-Q_{i\ell
k})\mathrm{TN}_+(0,\sigma,\mu) \\
&=& \mu+ \sigma\{Q_{i\ell k} \mathrm{TN}_+(0,1,-\mu/\sigma) -
(1-Q_{i\ell k})\mathrm{TN}_+(0,1,\mu/\sigma)\}.
\end{eqnarray*}
Generating $\mathrm{TN}_+(0,1,c)$ is easy: if $c < 0$, simply do
rejection sampling of a~$\Normal(0,1)$ until you get one that is $>
c$. If $c> 0$, there is an adaptive rejection scheme [Robert (\citeyear{Rob95})].

%s7.13 ###
\subsection{\texorpdfstring{Complete conditionals for $W_{i2k}$, $W_{i4k}$, $W_{i6k}$,
$W_{i8k}$, $W_{i,10,k}$ and $W_{i,12,k}$ when not observed.}{Complete conditionals for $W_{i2k}$, $W_{i4k}$, $W_{i6k}$,
$W_{i8k}$, $W_{i,10,k}$ and $W_{i,12,k}$ when not observed}}\label{a.6}

For $p = 2,4,6,8,10,12$, the variable $W_{ipk}$ is not observed when
$Q_{i,p -1,k} = 0$, or, equivalently, when $W_{i,p-1 ,k} < 0$. Except
for irrelevant constants,
\begin{eqnarray*}
\log[W_{ip k}|\rest] &=&
-\frac{1}{2} w_i \sum_j \sum_{\ell} \sigma_{\epsilon
}^{j\ell} (W_{ijk}-X_{ijk}\trans\beta_j-U_{ij}) (W_{i\ell
k}-X_{i\ell k}\trans\beta_\ell-U_{i\ell}) \\
&=& -\frac{1}{2}W_{ipk}^2 \mathcal{C}_2^{-1} + \mathcal{C}_1 W_{ipk},
\end{eqnarray*}
where, using the convention of Appendix \ref{subsec:ui},
\begin{eqnarray*}
\mathcal{C}_2 &=& 1 / (\sigma_{\epsilon}^{pp}); \\
\mathcal{C}_1 &=& \sigma_{\epsilon}^{pp}(X_{ipk}\trans\beta_p +
U_{ip}) - \sum_{\ell\neq p}\sigma_{\epsilon}^{p\ell
}(W_{i\ell k}-X_{i\ell k}\trans\beta_\ell-U_{i\ell}).
\end{eqnarray*}
Therefore,
\begin{eqnarray*}
[W_{ipk}|\rest] = Q_{ipk}Q_{i,p-1,k} + (1 - Q_{i,p-1,k})\Normal(\mathcal{C}_2\mathcal{C}_1,\mathcal{C}_2).
\end{eqnarray*}

%s7.14 ###
\subsection{\texorpdfstring{Usual intake, standardization and transformation.}{Usual intake, standardization and transformation}}\label
{backtransformation}

Here we present detailed formulas for functions defined in Section \ref
{subsec:dhei}. When $\lambda= 0$, the back-transformation is
\begin{eqnarray*}
g^{-1}_{\mathrm{tr}}(z,0) &=& \exp\bigl\{\mu(0) + \sigma(0) z / \sqrt{2} \bigr\};
\\
\partial^2 g^{-1}_{\mathrm{tr}}(z,0) / \partial z^2 &=& \frac{\sigma
^2(0)}{2} g^{-1}_{\mathrm{tr}}(z,0).
\end{eqnarray*}
When $\lambda\neq0$, the back-transformation is
\begin{eqnarray*}
g^{-1}_{\mathrm{tr}}(z,\lambda) &=& \bigl[1 + \lambda\bigl\{\mu( \lambda) +
\sigma( \lambda) z / \sqrt{2} \bigr\} \bigr]^{1/\lambda}; \\
\partial^2 g^{-1}_{\mathrm{tr}}(z,\lambda) / \partial z^2 &=& \frac
{\sigma^2(\lambda)}{2}(1-\lambda)
\bigl[1 + \lambda\bigl\{\mu(\lambda) + \sigma(\lambda) z / \sqrt{2} \bigr\}
\bigr]^{-2+1/\lambda}.
\end{eqnarray*}

%s7.15 ###
\subsection{\texorpdfstring{Transformation estimation.}{Transformation estimation}}\label{esttransformation}

As part of an earlier project [Freedman et~al. (\citeyear{Freetal10})], we estimated
the transformations for one food/nutrient at a time using the method of
\citet{Kipetal09}, both for the data and also for each BRR weighted
data set. To facilitate comparison with the one food/nutrient at a time
analysis, in our analysis of all HEI-2005 components, we used these
transformations as well. Of course, our methods can be generalized to
allow for estimation of the transformations as well. By allowing a~%
different transformation for each BRR weighted data set, we have
captured the variation due to estimation of the transformations.
\end{appendix}

%
%This paper forms part of Zhang's Ph.D. dissertation at Texas A\&M
%University. Zhang and Carroll's research was supported by a grant from
%the National Cancer Institute (CA57030). This work was also supported
%by National Science Foundation Instrumentation grant number 0922866.

%%\sname{
%Included in the supplementary materials [\citet{Zhaetal}] are (a)
%additional tables in a pdf file; (b) data files of the NHANES data used
%in the analysis; and (c) Matlab programs for the data analysis.
% %}
% \slink[url]{http://???/???}
% \sdescription{???}

\begin{supplement}[id=suppA]
\sname{Supplement A}
\stitle{Additional tables}
\slink[doi]{10.1214/10-AOAS446SUPPA} %[doi,text={...}] - jei reikia
%suskaldyti doi
\slink[url]{http://lib.stat.cmu.edu/aoas/446/supplementA.pdf}
\sdescription{}
\end{supplement}

\begin{supplement}[id=suppB]
\sname{Supplement B}
\stitle{Data files of the NHANES data used
in the analysis}
\slink[doi]{10.1214/10-AOAS446SUPPB} %[doi,text={...}] - jei reikia
%suskaldyti doi
\slink[url]{http://lib.stat.cmu.edu/aoas/446/supplementB.zip}
\sdatatype{.zip}
\end{supplement}

\begin{supplement}[id=suppC]
\sname{Supplement C}
\stitle{Matlab programs for the data analysis\\}
\slink[doi]{10.1214/10-AOAS446SUPPC} %[doi,text={...}] - jei reikia
%suskaldyti doi
\slink[url]{http://lib.stat.cmu.edu/aoas/446/supplementC.zip}
\sdatatype{.zip}
\end{supplement}

% imsref loaded by dianan, 2011-02-02 11:06:15
%

\printaddresses

\end{document}